\documentclass[aps,prx, reprint,longbibliography, nofootinbib,superscriptaddress]{revtex4-1}

\usepackage{graphicx,amsmath,amssymb,color,amsfonts,bm,physics}
\usepackage{mathtools,lipsum}
\usepackage[dvipsnames]{xcolor}
\usepackage[misc]{ifsym}

\usepackage[colorlinks=true]{hyperref}  
\hypersetup{
    bookmarks=true,         
    unicode=false,          
    pdftoolbar=true,        
    pdfmenubar=true,        
    pdffitwindow=false,     
    pdfstartview={FitH},    
    pdftitle={},    
    pdfauthor={},     
    pdfsubject={},   
    pdfcreator={},   
    pdfproducer={}, 
    pdfkeywords={} {} {}, 
    pdfnewwindow=true,      
    colorlinks=true,       
    linkcolor=blue, 
    citecolor=blue,        
    filecolor=magenta,      
    urlcolor=blue,           
        breaklinks=true
} 

\newcommand{\be}{\begin{equation}}
\newcommand{\ee}{\end{equation}}
\newcommand{\bea}{\begin{eqnarray}}
\newcommand{\eea}{\end{eqnarray}}

\begin{document}

\title{Quantifying quantum chaos through microcanonical distributions of entanglement}

\author{Joaquin F. Rodriguez-Nieva}
\thanks{Equal Contribution}
\affiliation{Department of Physics \& Astronomy, Texas A\&M University, College Station, TX 77843}
\email[Corresponding author: ]{jrodrigueznieva@tamu.edu}


\author{Cheryne Jonay}
\thanks{Equal Contribution}
\affiliation{Department of Physics, Stanford University, Stanford, CA 94305}

\author{Vedika Khemani}
\affiliation{Department of Physics, Stanford University, Stanford, CA 94305}

\date{\today}

\begin{abstract}
A characteristic feature of ``quantum chaotic" systems is that their eigenspectra and eigenstates display universal statistical properties described by random matrix theory (RMT). However, eigenstates of local systems also encode structure beyond RMT. To capture this,  we introduce a quantitative metric for quantum chaos which utilizes the Kullback–Leibler divergence to compare the microcanonical distribution of entanglement entropy (EE) of midspectrum eigenstates with a reference RMT distribution generated by pure random states (with appropriate constraints). The metric compares not just the averages of the distributions, but also higher moments.  The differences in moments are compared on a highly-resolved scale set by the standard deviation of the RMT distribution, which is exponentially small in system size. This distinguishes between chaotic and integrable behavior, and also quantifies the {\it degree} of chaos in systems assumed to be chaotic. We study this metric in local minimally structured Floquet random circuits, as well as a canonical family of many-body Hamiltonians, the mixed field Ising model (MFIM). For Hamiltonian systems, the reference random distribution must be constrained to incorporate the effect of energy conservation. The metric captures deviations from RMT across all models and parameters, including those that have been previously identified as strongly chaotic, and for which other diagnostics of chaos such as level spacing statistics look strongly thermal. In Floquet circuits, the dominant source of deviations is the second moment of the distribution, and this persists for all system sizes. For the MFIM, we find significant variation of the KL divergence in parameter space. Notably, we find a small region where deviations from RMT are minimized, suggesting that ``maximally chaotic" Hamiltonians may exist in fine-tuned pockets of parameter space.

\end{abstract}



\maketitle


\section{Introduction}
The emergence of statistical mechanics from the dynamics of isolated quantum systems is a topic of fundamental interest~\cite{1991PRL_Deutsch,1994PRE_Srednicki,2008Nature_Rigol,nandkishore_huse_annualreview}. While the foundations of this subject date back to the birth of quantum mechanics, the topic has seen a recent revival due to remarkable experimental advances in preparing isolated quantum systems that can be  coherently evolved over unprecedented time scales~\cite{2012NP_Bloch,2016Science_greiner,2018PRX_Lev,2016PRL_Schaetz,2018PRL_rydbergthermalization}. 
Unlike in classical systems, notions of ``ergodicity" and ``chaos" in many-body quantum systems are much more ill-defined.  One  prevailing approach to characterize quantum chaos is through the eigensystem properties of Hamiltonians (or time-evolution operators), specifically with respect to the emergence of universal behavior described 
by random matrix theory (RMT). This applies both to correlations of eigenvalues, such as the level spacing statistics \cite{atas_distribution_2013,atas_joint_2013,oganesyan_localization_2006} or the spectral form factor~\cite{Bertini_2018,Gharibyan_2018, chan2018solution, chan2018spectral, Friedman_2019, Garratt_2021}, and to the properties of eigenstates. In particular, the central conjecture underpinning the celebrated eigenstate thermalization hypothesis (ETH) is that highly-excited (infinite temperature) eigenstates of chaotic quantum systems look like random pure states within subsystems~\cite{1991PRL_Deutsch,1994PRE_Srednicki,2008Nature_Rigol,2018PRX_Grover, LuGrover, 
 2016AnnPhys_ETHreview,2018RPP_Deutsch,dymarsky_subsystem_2018}. This is reflected in the eigenstate expectation values of local observables \cite{1999JPA_Srednicki,dalessio_quantum_2016}, as well as the behavior of the eigenstate entanglement entropy (EE)~\cite{2017PRL_EEchaotic, Murthy_2019, 2018PRX_Grover, LuGrover, Bianchi-VidmarReview}. 

In recent years, a series of works have used the von Neumann entanglement entropy to refine the correspondence between midspectrum eigenstates of \emph{local}, \emph{physical}  Hamiltonians\footnote{We may henceforth drop these qualifiers (local, physical) but, throughout this work, we will only consider Hamiltonians with short-range, few-body interactions represented by sparse matrices. For Floquet systems, the interactions are time-dependent but still instantaeously local.} (or Floquet systems) and random pure states (see Ref.~\onlinecite{Bianchi-VidmarReview} for a recent review). A widely prevailing expectation~\cite{Bianchi-VidmarReview} is that --- in the absence of additional conservation laws --- infinite temperature eigenstates of chaotic Hamiltonians (or eigenstates of chaotic Floquet systems) are nearly maximally entangled, with EE following the Page equation~\cite{1993PRL_Page} derived for random pure states (Eq.~\ref{eq:Savg}). More recently, the Page result has been generalized to pure random states in various physically relevant \emph{constrained} settings, in which the Hilbert space does \emph{not} factor into a tensor product of the Hilbert spaces of subsystems~\cite{1991PRL_Deutsch, 2019PRD_BianchiDona, 2017PRL_EEchaotic, vidmar_entanglement_2017, morampudi_universal_2020}.  A notable example is systems with a local additive conserved charge, such as particle number, for which the Hilbert space is a direct sum of tensor products in different charge sectors. In this case, the typical EE for pure random states constrained to a given charge sector was recently derived analytically by Bianchi and Dona~\cite{2019PRD_BianchiDona}, and found in numerical studies to agree well with the eigenstate entropy of chaotic local Hamiltonians with particle-number conservation~\cite{Bianchi-VidmarReview}.  
 
The fact that the eigenenergies and eigenstates of a wide range of non-random, sparse Hamiltonians numerically display universal RMT correlations (derived for random, dense matrices) is  quite remarkable; understanding why this happens is a longstanding question in the study of quantum chaos. 
In particular, RMT ensembles, by design, have no spatial correlations, while local Hamiltonians do. Further, while a random many-body wavefunction in a system of size $L$ has exponentially many ($O(\exp L)$) random parameters, a local Hamiltonian or Floquet unitary is specified with just polynomially many ($O(L)$) parameters. 

Thus, various recent papers have focused on the question of how the eigenspectra of local systems encode structure beyond the leading-order RMT behavior, even in the absence of additional symmetries or constraints. For example, Refs.~\cite{2022PRE_deviationsfromETH,huang_hamentanglement, huang_midspectrum, huang_deviation_2022,2023arxiv_rigol} have numerically and analytically studied systematic deviations between the EE of midspectrum Hamiltonian eigenstates and the Page entropy. Separately, Refs.~\cite{Dymarsky_deviation, 2022PRL_dymarsky, 2020PRE_beyondETH,2021PRE_eth_otocs, 2019PRE_kurchan,2019PRL_Chalker, Garratt_2021} showed that  matrix elements of local operators evaluated in the eigenbasis of Hamiltonian or Floquet systems are correlated  up to certain energy scales (or inverse time scales related to, but possibly parameterically larger than, the so-called Thouless time); these works clarified that these correlations can be understood through the existence of a ``light cone" in the growth of out-of-time ordered commutators in spatially local extended systems. In fact, eigenstates of local Hamiltonians ``know'' that they are \textit{not} RMT: the correlations encoded in a single eigenstate suffice to reconstruct the entire Hamiltonian~\cite{2018PRX_Grover, 2019Q_Ranard}. However, despite these various works, a systematic and unified understanding of eigenstate diagnostics of chaos, including universal deviations from RMT, is considerably less developed. This is in contrast to eigenvalue diagnostics of chaos for which analytic results for the spectral form factor have been derived for various systems, and have been shown to display RMT `ramp' behavior for times larger than the so-called Thouless time~\cite{Bertini_2018,Gharibyan_2018, chan2018solution, chan2018spectral, Friedman_2019, Garratt_2021}. The Thouless time encodes the effects of locality and, in general, grows with system size; the Thouless time is minimized (and system-size independent) in certain ``maximally chaotic" kicked-Ising Floquet models which are dual-unitary~\cite{Bertini_2018, 2019PRX_dualunitary, Gopalakrishnan_Lamacraft}.

\begin{figure}
\centering\includegraphics[scale = 1.0]{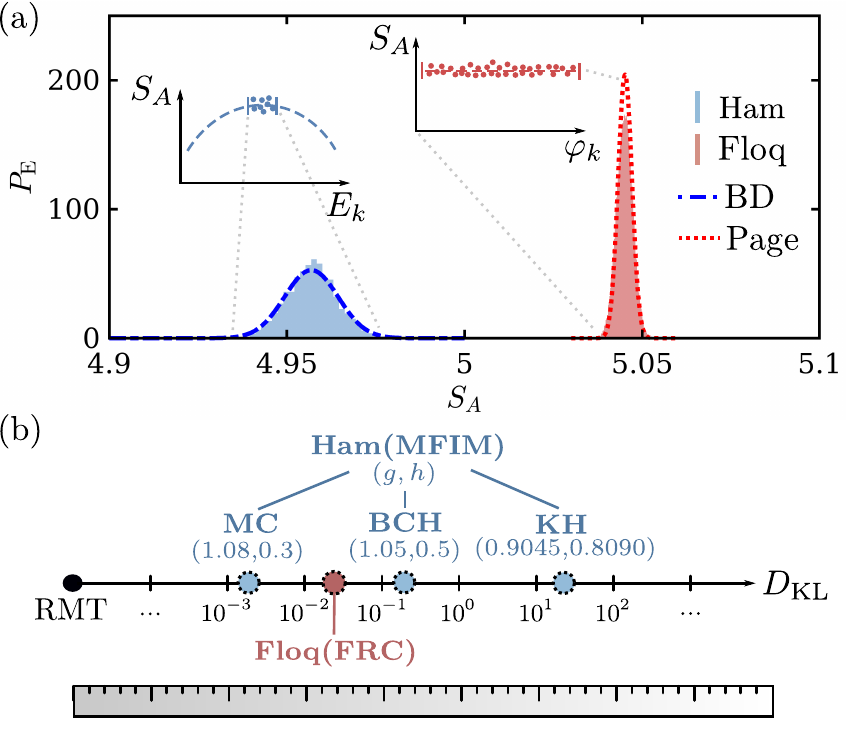}
\caption{\label{fig:midspectrum}
(a) Histograms of EE of (left) midspectrum eigenstates of the Mixed Field Ising model (MFIM) Hamiltonian, see Eq.(\ref{eq:TFIM}),  and (right) eigenstates of a Floquet random circuit (FRC) with nearest-neighbor Haar random gates, see Eq.~(\ref{eq:U4}). The MFIM parameters are chosen to be strongly chaotic, with  $g = 1.08$ and $h = 0.3$; FRCs are averaged over 50 circuit realizations.  For comparison, we plot the reference RMT distributions, Page (dotted lines) and Bianchi-Dona (BD; dashed-dotted lines), see Secs.~\ref{sec:page} and ~\ref{sec:BD}. We see that the former is the appropriate reference distribution for Floquet systems, while the latter describes Hamiltonians with energy conservation.  The reference distributions are plotted as Gaussian functions with analytically-known means and standard deviations (see App.~\ref{app:exactmoments}). For all distributions, we consider a system of size $L = 16$ and a subsystem of size $L_A = 8$. (b) The distance between microcanonical distribution of eigenstate EE and RMT, as quantified through the Kullback-Liebler (KL) divergence $D_{\rm KL}$,  Eq.~(\ref{eq:DKL}). Shown are the values of $D_{\rm KL}$ plotted for a system of size $L=16$ for the FRC and different MFIM parameters studied in the literature or discussed later in the main text: Maximally Chaotic (MC), Banulus-Cirac-Hastings (BCH)~\cite{BanulsCiracHastings}, and Kim-Huse (KH)~\cite{Kim_2013}, see Fig.~\ref{fig:DKL}. 
}
\end{figure}

Our perspective in this work is to introduce an eigenstate metric to quantify how chaotic a many-body Hamiltonian/Floquet system is --- in the sense of providing a continuous ``ruler" measuring deviations from RMT (Fig.~\ref{fig:midspectrum}). The metric that we propose measures a ``distance" to the appropriate RMT ensemble by computing the Kullback–Leibler (KL) divergence, $D_{\rm KL}(P_{\rm E}, P_{\rm R})$, between the microcanonical distribution of eigenstate EE, $P_{\rm E}(S_A)$, and an appropriate reference RMT distribution, $P_{\rm R}(S_A)$. Here $S_A$ refers to the EE of a subsystem $A$ of a pure state, which is chosen to be either an eigenstate  or a random state. In other words, we ask how well the microcanonical ensemble of eigenstates reproduces the distribution of EE generated by an (appropriately constrained) ensemble of random pure states. For Hamiltonian systems, the microcanonical ensemble is obtained from a narrow window of eigenstates centered at infinite temperature, while we use the entire eigenspectrum for Floquet systems. 
The measure goes beyond the first moment by also incorporating higher moments of the distributions, which depend on the fluctuations of EE across eigenstates. 
In fact, we will show that these fluctuations play a key role, and capture deviations from RMT that may not be visible in the first moment in certain systems. 

\subsection{Summary of Results}

We study two models in this work: local Floquet random circuits (FRCs), and a family of mixed-field Ising model (MFIM) Hamiltonians parameterized by the strength of the transverse and longitudial fields. Our main results are summarized as follows: 

First, in minimally structured FRCs (which only retain locality, and have no other structure or conservation laws), we find that the dominant contribution to $D_{\rm KL}$ comes from the microcanonical \emph{fluctuations} of EE. In particular, we find that $\sigma_{\rm E}$, the standard deviation of the microcanonical distribution, is systematically larger than $\sigma_{\rm R}$, the standard deviation of the Page distribution, which is the reference RMT distribution. In contrast, the means, $\mu_{\rm E}$ and $\mu_{\rm R}$, are much better converged. We note that $D_{\rm KL}$ furnishes a very finely resolved comparison between the microcanonical and RMT distributions by normalizing the differences between the distributions by $\sigma_{\rm R}$, so that the contributions of the first two moments of the microcanonical distribution to $D_{\rm KL}$ are functions of the ratios $|\mu_{\rm E}-\mu_{\rm R}|/\sigma_{\rm R}$ and $\sigma_{\rm E}/\sigma_{\rm R}$, see Eq.~\eqref{eq:DKL_approx}. For the reference RMT distribution, the standard deviation is exponentially small in system size: $\sigma_{\rm R}\sim \sqrt{2^{-L}}$.   Thus, unlike prior works, $D_{\rm KL}$ measures not just whether $|\mu_{\rm E}-\mu_{\rm R}|$ decreases with system size, 
or whether $\sigma_{\rm E}$ displays exponentially small scaling similar to RMT, but rather \emph{it probes these differences on the exponentially small scale set by $\sigma_{\rm R}$}. For FRCs, we find that the difference in mean between the microcanonical and RMT distributions is small even on the scale of $\sigma_{\rm R}$, while the ratio $\sigma_{\rm E}/\sigma_{\rm R}$ shows a sizable and positive $O(1)$ departure from 1 which appears stable with system size (Fig.~\ref{fig:FRC}). In other words, for FRCs, we find that $\sigma_E \sim \sqrt{2^{-L}}$, but with a systematically larger prefactor than the reference Page distribution. We attribute this relative increase in $\sigma_{\rm E}$ to locality, since FRCs do not have any structure or symmetries beyond locality and time-periodicity. 

Second, even for Hamiltonian systems, we find that the difference between the microcanonical and RMT prediction is \emph{exponentially} small in system size (and comparable to $\sigma_{\rm R}$ in large regions of parameter space, away from integrability), \textit{provided} we suitably constrain the RMT ensemble. This may seem at odds with various works which recently noted that the EE of midspectrum Hamiltonian eigenstates shows a small but systematic $O(1)$ negative departure from the Page value~\cite{2022PRE_deviationsfromETH, huang_hamentanglement, huang_midspectrum, huang_deviation_2022,2023arxiv_rigol}. Remarkably, we show instead that the departure is captured if we instead compare $\mu_{\rm E}$ to the mean of the Bianchi-Dona (BD) distribution~\cite{2019PRD_BianchiDona} obtained for systems with a local $U(1)$ charge. In other words, a ``better" RMT ensemble for describing midspectrum eigenstates of local Hamiltonians is the BD distribution (as opposed to the Page distribution): energy conservation plays the role of an additive local charge, even at infinite temperature and in the absence of additional symmetries such as particle number conservation, and the BD distribution incorporates this important feature (see Fig.~\ref{fig:midspectrum}). This result is of independent interest, and updates numerous prior studies in which midspectrum eigenstates of local Hamiltonians (without additional symmetries) have been compared to the Page distribution~\cite{Bianchi-VidmarReview}. We note that the deviation between the average microcanonical EE for Hamiltonian systems and the Page entropy was also previously argued for by Huang in  Ref.~\cite{huang_hamentanglement, huang_midspectrum}, and agrees with the results obtained by using the BD ensemble. In Fig.~\ref{fig:midspectrum}, we plot the microcanonical EE distribution for a non-integrable Hamiltonian and Floquet random circuit, and show that they are well described (to leading order) by the BD and Page distributions respectively. In particular, Figure~\ref{fig:midspectrum} shows that the means of the two reference RMT distributions differ by  $\sim 0.1$, which entirely captures the $O(1)$ deviation from Page that has previously been observed for the EE of Hamiltonian eigenstates~\cite{2022PRE_deviationsfromETH, huang_hamentanglement, huang_midspectrum}.

Third, for Hamiltonian systems,  we find that $D_{\rm KL}$  shows significant variation in parameter space for the models and system sizes we study --- even in parameter regimes where other metrics of chaos such as level statistics have saturated to the RMT predictions (Fig.~\ref{fig:DKL}). 
Remarkably, there are small islands in parameter space that minimize $D_{\rm KL}$, and these are quite far in parameter space from ``standard" reference values that are widely used in studies of chaos in the MFIM, such as the ``Kim-Huse" (KH) parameters of Ref.~\cite{Kim_2013}.  At the strongly chaotic points, the deviations in both $|\mu_{\rm E}-\mu_{\rm R}|/\sigma_{\rm R}$ and $\sigma_{\rm E}/\sigma_{\rm R}$ are comparable, while these differences steeply increase 
away from the maximally chaotic regions, resulting in a large $D_{\rm KL}$. As such, our FRC and Hamiltonian results show that $D_{\rm KL}$ can resolve differences in microcanonical and RMT distributions, even if the moments of the former are ``exponentially close" to the latter. This variation in parameter space suggests that there might be ``maximally chaotic" Hamiltonians in judiciously tuned regions of parameter space, similar to maximally chaotic dual-unitary Floquet circuits~\cite{Bertini_2018, 2019PRX_dualunitary, Gopalakrishnan_Lamacraft} or minimally chaotic integrable models,  with the degree of chaos that is attainable being constrained by features such as locality or the type of allowed interactions. 

More generally, the approach to compare microcanonical and RMT distributions, including higher moments, is reminiscent of (but different in detail from) studies of unitary- and state- design  in quantum information theory~\cite{hunter-jones_unitary_2019,brandao_models_2021,cotler_emergent_2023,choi_emergent_2023,ho_exact_2022,ippoliti_solvable_2022}. For example, studies of $k-$design formation compare the lowest $k$ moments of candidate probability distributions over a unitary group against the uniform Haar distribution; this approach is particularly informative in understanding distributions for which the lower moments agree with the Haar distribution while higher moments show deviations. Likewise, a notable feature of our work is that the microcanonical standard deviation of EE is informative in characterizing chaos even in models where the mean agrees with RMT predictions. 

There are also practical advantages of quantifying chaos through the microcanonical statistics of entanglement entropy. A single Hamiltonian produces its own microcanonical ensemble of eigenstates which is then used to characterize chaos. This contrasts with other metrics like the spectral form factor which requires sampling over Hamiltonian ensembles~\cite{prange1997spectral}. It also makes the approach quantitative as it allows one to compare the degree of chaos for two different systems using the same well-characterized benchmark. Further,  only a relatively small number of eigenstates are required to characterize the distribution of EE up to the second moment, therefore making the method inexpensive and relatively easy to implement for relatively large system sizes through various shift-invert or polynomial filtering techniques for targeting eigenstates in small energy or quasienergy windows~\cite{2018SciPost_shiftinvert,Luitz_2021}. Finally, our metric of chaos is intrinsic to eigenstates and is therefore operator-independent, unlike methods that rely on susceptibility metrics which require specifying a perturbing operator~\cite{2020PRX_Sels_chaos}. 

The outline of the rest of this paper is as follows. In Sec.~\ref{sec:metric}, we introduce our metric for quantum chaos and describe its behavior in simple limits. In Sec.~\ref{sec:floquet}, we begin the discussion by introducing the relevant RMT distribution that will be used to quantify chaos in Floquet Random Circuits, namely, the Page distribution. We then present numerical results for the distribution of EE of eigenstates in minimally-structured Floquet random circuits, and discuss contributions of the first two moments to the distance measure. In Sec.~{\ref{sec:numerics}}, we begin by introducing the relevant RMT distribution that will be used to quantify chaos in Hamiltonian systems, namely, the BD distribution. We then present numerical results for the mixed field Ising model (MFIM), a paradigmatic Hamiltonian system that exhibits both integrable and chaotic regimes. We discuss the behavior of the eigenstate EE distribution in the proximity of maximally chaotic parameters, and the differences with respect to other metrics of chaos employing spectral statistics.  Finally, in Sec.~\ref{sec:discussion}, we summarize the main results of our work and discuss directions for future work.

\section{Quantifying quantum chaos}
\label{sec:metric}

Our goal is to quantify the degree of quantum chaos in a given Hamiltonian or Floquet unitary by comparing the microcanonical distribution of eigenstate EE, $P_{\rm E}(S_A)$, with a reference distribution for the EE of (appropriately constrained) pure random states, $P_{\rm R}(S_A)$. 
We will consider systems of size $L$ and Hilbert space dimension $d$ partitioned into two subsystems  $A$ and $B$ with sizes $L_A \leq L_B$, respectively. We will find it convenient to introduce the ratio $f=L_A/L \leq 1/2$.  All the numerical data in this work will be for one-dimensional spin 1/2 systems, but the methods readily generalize to higher dimensions and systems of qudits. 

To obtain the microcanonical distribution, we choose eigenstates $|k\rangle$ of the Hamiltonian/Floquet system with energy/quasienergy $E_k/\varphi_k$ respectively. The reduced density matrix obtained from eigenstate $|k\rangle$ is 
\be
\rho_{A,k} = {\rm Tr}_B [\rho_k],\quad\rho_k= |k\rangle\langle k|, 
\ee 
with associated von Neumann entanglement entropy 
\be
S_{A,k} = -{\rm Tr}\left[ \rho_{A,k} \log \rho_{A,k}\right] \label{SVN}.
\ee
For Hamiltonian systems, we construct the microcanonical distribution $P_{\rm E}(S_A)$ by computing the EE of eigenstates in a small window centered around the middle of the spectrum, {\it i.e.}, at an energy density corresponding to infinite temperature (details are discussed in Sec.\,\ref{sec:numerics}). For Floquet unitaries, there is no conserved energy or notion of temperature (or,  colloquially, all states are at infinite temperature);  therefore, the full eigenbasis corresponding to states at all quasienergies are used to construct $P_{\rm E}(S_A)$. 

Our goal is to define a distance between the microcanonical distribution $P_{\rm E}(S_A)$ and a reference random distribution $P_{\rm R}(S_A)$. A natural choice for the distance between distributions is the Kullback-Leibler (KL) divergence, 
\be
D_{\rm KL}(P_{\rm E}, P_{\rm R}) =  \int d{S_A} P_{\rm E}(S_A) \log \frac{P_{\rm E}(S_A)}{P_{\rm R}(S_A)} \ge 0.
\label{eq:DKL}
\ee
The KL divergence is the expectation of the logarithmic difference between probability distributions, and is a measure of the information loss when the reference distribution $P_R(S_A)$ is used to approximate the empiricial eigenstate distribution $P_E(S_A)$.  The KL divergence is a type of distance, since it is always non-negative and takes value 0 when the reference and empirical distribution are equal. However, it is not a metric distance because it is asymmetric in the two distributions and does not satisfy the triangle inequality.

Let us evaluate Eq.(\ref{eq:DKL}) in a simple yet important limit. Given the first two moments $\mu_{\rm E/R}$ and $\sigma_{\rm E/R}$ of the empirical (or eigenstate/microcanonical) and reference (or random state) distributions,  we make a Gaussian approximation for both $P_{\rm E}(S_A) \approx \frac{1}{\sqrt{2\pi\sigma_E^2}}{\rm exp}\left[-\frac{(S_A-\mu_{\rm E})^2}{2\sigma_{\rm E}^2}\right]$ and $P_{\rm R}(S_A) \approx \frac{1}{\sqrt{2\pi\sigma_{\rm R}^2}}{\rm exp}\left[-\frac{(S_A-\mu_{\rm R})^2}{2\sigma_{\rm R}^2}\right]$. Within this approximation, $D_{\rm KL}$ quantifies the difference between means (relative to $\sigma_{\rm R}$) and the ratio $\frac{\sigma_{\rm E}}{\sigma_{\rm R}}$ through the non-linear relations
\begin{align}
D_{\rm KL} &= D_{\rm KL}^{(1)} + D_{\rm KL}^{(2)}, \nonumber \\
D_{\rm KL}^{(1)} &= \frac{(\mu_{\rm E}-\mu_{\rm R})^2}{2\sigma_{\rm R}^2}, \nonumber \\
D_{\rm KL}^{(2)} &=\frac{1}{2}\left[\left(\frac{\sigma_{\rm E}}{\sigma_{\rm R}}\right)^2 -1\right] - \log \frac{\sigma_{\rm E}}{\sigma_{\rm R}}.
\label{eq:DKL_approx}
\end{align}
In what follows, we will make this Gaussian approximation and only focus on the first two moments for the reference and empirical distributions while computing $D_{\rm KL}$. While this isn't strictly accurate (the strict upper bound on the value of entropy produces a skewness, for instance, which has been computed for random states with and without charge conservation symmetry~\cite{2019PRD_BianchiDona}), it is still a good approximation because the higher moments scale with increasing powers of $1/d$, where $d$ is the Hilbert space dimension~\cite{2019PRD_BianchiDona}. 

The different $P_{\rm R}(S_A)$ studied in this work, {\it i.e.}, the Page and Bianchi-Dona distributions, will be discussed in the following sections. Both distributions have $\sigma_{\rm R} \sim \sqrt{2^{-L}}$; thus, as mentioned earlier, $D_{\rm KL}$ provides a highly resolved comparison between the microcanonical and reference RMT distributions by comparing the differences between their moments on the exponentially small scale set by $\sigma_R$, as seen by the expressions for  $D_{\rm KL}^{(1,2)}$ in Eq.~\eqref{eq:DKL_approx}.

\section{Floquet systems}
\label{sec:floquet}

We start by analyzing a minimally-structured model of chaotic thermalizing dynamics, namely, Floquet random circuits that only feature locality and no other conservation laws.  In the absence of energy or U(1) conservation, we employ the Page distribution as the reference RMT distribution. We include a discussion of $\mu_{\rm R}, \sigma_{\rm R}$ for the Page distribution for completeness and to set notation, before presenting our numerical results.

\subsection{Reference distribution I: the Page distribution}
\label{sec:page}

We begin by recapitulating the `Page Distribution' for the bipartite entanglement entropy of pure random states chosen uniformly with respect to the Haar measure,  unconstrained by symmetry. Our nomenclature takes some historical liberties: in fact, Page only conjectured (and partially proved) the expression for the first moment of this distribution~\cite{1993PRL_Page}; explicit derivations for the first and higher moments were later furnished in Refs.~\cite{2016PRE_entanglementdispersion,2017PRE_entanglementvariance_proof, 2019PRD_BianchiDona}. 

As mentioned above, we will focus on the first two moments of the Page distribution in this work. The exact analytical expressions for these (for finite system sizes) are reproduced in Appendix~\ref{app:exactmoments}, and used in our numerical comparisons below. We briefly discuss these in the limits $L_A,L_B \gg 1$ and when $0<f\leq 0.5$ is a finite fraction as $L \rightarrow \infty$. The first moment in this limit is approximated as:
\be
\langle S_A \rangle_{\rm P} \equiv \mu_{\rm P}(f) \approx \log(d_A) -\frac{d_A}{2d_B},
\ee
where $d_A, d_B$ refer to the Hilbert space dimensions of subsystems $A,B$ respectively. We 
will henceforth use the subscript `P' for `Page' in order to denote moments computed with respect to the Haar measure. For a system of qubits, this reduces to
\be
\mu_{\rm P}(f) \approx fL \log(2)-2^{-L(1-2f)-1}.
\label{eq:Savg}
\ee
The first term is the volume law term which describes 
an entanglement entropy scaling with the size of subsystem $A$, $L_A = fL$, while the second term is the `Page correction' which is exponentially small when $f < 1/2$. For $f=1/2$, it gives rise to a `half-bit' shift: 
\be
\mu_{\rm P}\left(f= \frac{1}{2}\right) \approx \frac{L}{2} \log(2)-\frac{1}{2}.
\ee
More recently, the second moment of the distribution was calculated using various techniques~\cite{2016PRE_entanglementdispersion,2017PRE_entanglementvariance_proof,2019PRD_BianchiDona} and, in the limit $L_A,L_B \gg 1$ is approximated as~\cite{Bianchi-VidmarReview}
\be
\sigma_{\rm P}^2 (f)\approx \left(\frac{1}{2}-\frac{1}{4}\delta_{f,\frac{1}{2}}\right)\frac{1}{d_B^2} = \left(\frac{1}{2}-\frac{1}{4}\delta_{f,\frac{1}{2}}\right)  2^{-2L(1-f)}.
\label{eq:Svar}
\ee
 The second moment of the distribution $\sigma_{\rm P}$ is exponentially small in $L$, and scales as $\sqrt{1/d}$ at $f=1/2$:
 \be
\sigma_{\rm P} \left(f=\frac{1}{2}\right)\approx \frac{1}{2} \sqrt{ 2^{-L}}.
\ee
Because $\sigma_{\rm P}/ \mu_{\rm P} \ll 1$, a \emph{typical} random state will have EE given by Eq.(\ref{eq:Savg}).

\subsection{Numerical Results}

 We consider a Floquet random unitary circuit comprised of two layers of Haar random two-site unitary gates acting on odd and even bonds, $U_F = U_{\rm odd}U_{\rm even}$, in a spin 1/2 chain of even length $L$ with periodic boundary conditions:
\begin{align}
    U_{\rm even}& = U_{0,1} \otimes U_{2,3} \otimes U_{4,5} \cdots \otimes U_{L-2, L-1},\nonumber \\
    U_{\rm odd} & = U_{1,2} \otimes U_{3,4} \otimes U_{5,6} \cdots \otimes U_{L-1, 0}. 
\label{eq:U4}
\end{align}
Each of the unitary matrices $U_{i,i+1}$ are chosen randomly and uniformly from the Haar measure on $U(4)$. The time-evolution operator for integer times $t$ is $U(t) = U_F^t$. We emphasize that the system has locality and time-periodicity, but no additional symmetries. 

The microcanonical EE distribution $P_E(S_A)$ is computed as a function of $L$ for $L_A = L/2$, and compared with the Page distribution of EE for pure random states. The mean and standard deviation of the EE distribution are denoted $\mu_U, \sigma_U$ respectively, where we use the subscript `$U$' to denote eigenstates of unitary circuits, which is more descriptive than the $\mu_{\rm E}$, $\sigma_{\rm E}$ symbols introduced in Sec.~\ref{sec:metric} (which referred collectively to moments of eigenstate distributions of either Hamiltonian or Floquet systems). For each system size, we average the microcanonical EE distribution over 50 circuit realizations, and compute the microcanonical distribution from the entire Floquet spectrum for a given realization.  For $L\le 14$, we use exact diagonalization to compute all eigenvectors and quasienergies for $U_F$. For $L=16$, we use the polynomial filtering diagonalization method introduced in~\cite{Luitz_2021} to obtain a total of 2000 eigenstates per circuit. These are obtained in groups of 50 states centered around 40 evenly spaced quasienergies distributed across the full quasienergy spectrum. In {Appendix} \ref{app:windowsize}, we show that our results are not sensitive to whether the microcanonical distribution is obtained using the full spectrum vs. narrower windows of states clustered around particular quasienergies.

\begin{figure}
\centering\includegraphics[scale = 1.0]{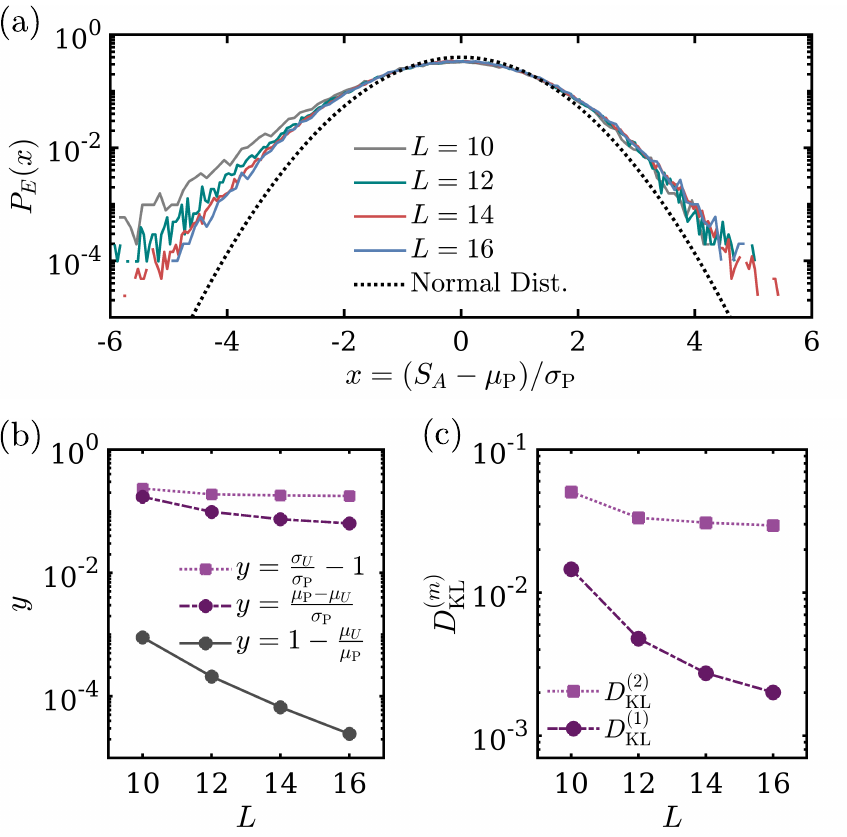}
\caption{(a) Histogram of eigenstate EE for a  FRC as a function of system size showing a broader distribution than that predicted by RMT, see also Fig.\ref{fig:midspectrum}. Shown with dotted lines is the standard normal distirbution. (b) System size scaling of the first two moments of the EE distribution relative to RMT behavior: shown are convergence of difference in EE means (plotted both relative to $\mu_{\rm P}$ and to $\sigma_{\rm P}$) while finite deviations of standard deviations persists for all system sizes. (c) Contributions to $D_{\rm KL}$ by the first two moments of the EE distribution of FRC eigenstates.}
\label{fig:FRC}
\end{figure}

We showed in Fig.~\ref{fig:midspectrum}(a) that the distribution $P_{\rm E}(S_A)$ for $L=16$ and $L_A=8$ shows good agreement with the Page distribution, so that the entanglement properties of eigenstates of FRCs are well described by the entanglement properties of Haar random states, as expected from numerous prior works.  We now  provide a more finely resolved comparison of the microcanonical and Page distributions for different system sizes; in particular, we compare both the first and second moments of these distributions, and probe differences on the exponentially small scale set by $\sigma_{\rm P}$. In order to do this, for each eigenstate, we shift the EE of the eigenstate by $\mu_{\rm P}$ and normalize by $\sigma_{\rm P}$ (the expressions  for $\mu_{\rm P}$ and $\sigma_{\rm P}$ depend on $L$ and are provided in Eqs.~\eqref{SAvgExact}, \eqref{SVarExact}). With these transformations, the  Page distribution (within the Gaussian approximation) reduces to a standard normal distribution for all $L$. The shifted and rescaled distributions for $x = (S_A - \mu_{\rm P})/\sigma_{\rm P}$ are plotted in Fig.\ref{fig:FRC}(a), with the standard normal distribution shown for comparison. We see from this figure that while the means of the microcanonical and Page distributions are in close agreement, we find that $P_{\rm E}(S_A)$ converges with increasing $L$ to a \emph{wider} distribution than Page, {\it i.e.}, with larger standard deviation. Both the right and left tails of $P_{\rm E}(S_A)$ contribute to the increased width so that, in comparison to the Page distribution over random states, it is more likely for the entropy of FRC eigenstates to show larger positive \emph{and} negative deviations from the Page mean.  

In Fig.~\ref{fig:FRC}(b), we compare the first two moments of the Page and microcanonical distributions. We find that $\mu_{\rm P} > \mu_U$, but the difference $(\mu_P-\mu_U)$ converges to zero exponentially with $L$. First, we normalize $(\mu_{\rm P}-\mu_U)$ by $\mu_{\rm P}\sim L$, and notice the exponential decrease in  $(\mu_{\rm P}-\mu_U)/\mu_{\rm P}$ with increasing $L$. Next, even upon normalizing by $\sigma_{\rm P} \sim\sqrt{2^{-L}}$, which is itself exponentially decreasing with $L$, we notice that not only is  $(\mu_{\rm P}-\mu_{U})$ an order or magnitude smaller than $\sigma_{\rm P}$, but that the ratio $(\mu_{\rm P}-\mu_U)/\sigma_{\rm P}$ still shows a weak decrease with increasing $L$ (consistent with plateauing at larger $L$). In contrast, the differences in standard deviations is more stark:  the ratio of standard deviations $\sigma_U/\sigma_{\rm P}$ plateaus to a constant value $\sim 1.2$, so that the microcanonical standard deviation is about $\sim 20\%$ larger than the Page standard deviation. In other words, while $\sigma_U \sim \sqrt{2^{-L}}$ shows the same exponential scaling as $\sigma_{\rm P}$, the prefactor for the scaling is $\sim 20\%$ larger. 

The different moments (normalized by $\sigma_{\rm P}$) contribute to $D_{\rm KL}$ according to Eq.~(\ref{eq:DKL_approx}). We show in Fig.~\ref{fig:FRC}(c) that $D_{\rm KL}^{(2)}$, the contribution from the second moment, is more than an order of magnitude larger than $D_{\rm KL}^{(1)}$, the contribution from the first moment.  In other words, while the average entropy of eigenstates of FRCs is well described by the average entropy of random pure states, the fluctuations of the microcanonical distribution are markedly larger, so that an increase in standard deviation is the dominant source of difference between the microcanonical and Page distributions. Since the FRC lacks any structure beyond locality, we attribute the relative increase in the standard deviation of $P_E(S_A)$ as a correction to RMT that arises from locality\footnote{As a side remark, we note that the orthogonality of eigenstates is not responsible for the deviation from RMT, as discussed in~\cite{2022PRE_deviationsfromETH}. Indeed, we checked that the distribution of entanglement entropy produced by a global $U(N)$ random unitary ({\it i.e.}, in the absence of locality)  does not exhibit deviations from the Page distribution for the first two moments.}. Indeed, in {Appendix}~\ref{app:gaterange} we show increasing convergence to the Page moments as the constraint of locality is relaxed by increasing the period of the Floquet circuit (i.e. by incorporating increasing numbers of even-odd layers). We obtained similar results by increasing the gate range while keeping the period fixed ({Appendix}~\ref{app:gaterange}).

There are (at least) two mechanisms by which locality could affect entropy fluctuations. First, the entanglement entropy of eigenstates of local FRCs will be sensitive to the entangling properties of the local unitary gates straddling the entanglement cuts (in our case, these are two two-site gates displaced by half the system size). These local unitary gates have a much larger likelihood of being either weakly entangling (i.e. close to the identity) or maximally entangling (for example, close to the iSWAP gate) in comparison to global Haar random unitaries. Thus, even though the microcanonical distribution of entropy for local FRCs shows the same exponential scaling as for global Haar random circuits, $\sigma_U\sim \sqrt{2^{-L}}$, the effect of locality (and, in particular, the distribution in entangling power for local gates) could contribute to a larger prefactor for $\sigma_U$. This suggests that exploring different families of local FRCs could allow us to tune $\sigma_U$ and $D_{\rm KL}$ to identify `maximally chaotic' families of circuits where $D_{\rm KL}$ is minimized. This would be an eigenstate analog of the property of `maximally chaotic' dual-unitary circuits for which the Thouless time in the spectral form factor is minimized (and system size independent) despite spatial locality~\cite{Bertini_2018}. We defer this analysis to future work. 

Second, Refs.~\cite{2021PRE_eth_otocs, 2019PRE_kurchan,2019PRL_Chalker, Garratt_2021} showed that there are microcanonical correlations between expectation values of local operators computed in eigenstates of local Floquet circuits (which also imply correlations between reduced density matrices of eigenstates, which are also used to compute entanglement entropy). These correlations are not present in an RMT description of the system and they
arise from the presence of light-cones in the spreading of local operators (or scrambling of quantum information) in spatially local systems. In particular, these correlations will be present even in dual-unitary models for which the Thouless time is minimized and agrees with RMT. While this analysis does not directly apply to the microcanonical distribution of half-system von-Neumann entanglement entropy, it is reasonable to expect that eigenstate correlations of reduced density matrices could also affect the eigenstate entropy distribution. Understanding this connection better, and teasing apart different effects induced by locality that may contribute  to increased $\sigma_U$ is also an interesting direction for future work. 

In sum, the results of this section corroborate that $D_{\rm KL}$, particularly microcanonical entropy fluctuations, furnishes a sensitive and easy-to-characterize metric for quantifying chaos via deviations from RMT. This metric can be used to compare different models with ease, and can encapsulate various different effects of locality that may individually be more difficult to calculate, benchmark and compare across models and observables.   
We now turn to studying this metric in a family of mixed-field Ising Hamiltonians parameterized by the strength of a transverse and longitudinal field. We will identify maximally chaotic models within this family of Hamiltonians by minimizing $D_{\rm KL}$ in parameter space.

\section{Hamiltonian systems}
\label{sec:numerics}

We now consider the microcanonical distribution of EE produced by midspectrum eigenstates in the Mixed Field Ising model (MFIM), a paradigmatic model that exhibits both chaotic and integrable limits depending on the model parameters. We first present the Bianchi-Dona distribution and argue why it serves as a better reference random distribution (as compared to the Page distribution) for Hamiltonian systems with energy conservation. We then introduce the MFIM model, generate the empirical distribution of midspectrum eigenstate entropies,  $P_{\rm E}(S_A)$, and compare the first two moments with the BD distribution. We quantity the distance between the distributions using the KL divergence, Eq.~(\ref{eq:DKL_approx}), and compare with conventional measures of quantum chaos such as level statistics. 

\subsection{Reference distribution II: the Bianchi-Dona distribution}
\label{sec:BD}

The presence of symmetries can affect the distribution of the EE.  We will argue that the symmmetry of primary interest to this work corresponds to the conservation of an additive local scalar charge $M$. We will refer to the distribution of entanglement entropy for pure random states subject to this constraint as the Bianchi-Dona (BD) distribution~\cite{2019PRD_BianchiDona}, but note that some aspects of this ensemble of random states were previously also discussed by Huang in~\cite{huang_hamentanglement, huang_midspectrum}. We first describe this distribution, and then argue why it captures important contributions of energy conservation to the EE of midspectrum eigenstates of spatially local Hamiltonians. 

BD considered systems with an additive local charge which decomposes between a bipartion of the system into subsystems $A$ and $B$ as $M = M_A + M_B$.  For concreteness, it is convenient to think of $0 \leq M \leq L$ as an integer particle number, with each site only able to accommodate a maximum of one particle.  The Hilbert space ${\mathcal{H}(M)}$ of states with fixed charge $M$ no longer has a tensor product structure, but instead decomposes as a direct sum of tensor products: 
\be 
{\cal H}(M) = \bigoplus_{M_A ={\rm min}(0, M-L_B)}^{{\rm max}(M, L_A)} {\cal H}_A(M_A)\otimes{\cal H}_B(M-M_A).
\label{eq:factorization}
\ee
The Hilbert space dimension of ${\cal H}_A(M_A)$ is 
$d_{A,M_A} = \binom{L_A}{M_A}$, and the Hilbert space dimension of ${\cal H}_{\rm B}(M-M_A)$ is $d_{B,M-M_A}= \binom{L-L_A}{M-M_A}$. The total Hilbert space dimension is $\sum_{M_A} d_{A,M_A} d_{B,M-M_A} =d_M =\binom{L}{M}$. A random state with fixed total charge $|\Psi_M\rangle \in {\cal H}(M)$ can be expressed as a superposition of orthonormal basis states, 
\begin{equation}
    \ket{\Psi_M}=\sum_{M_A} \sum_{\alpha=1}^{d_{A, M_A}} \sum_{\beta=1}^{d_{B, M-M_A}} \psi^{(M_A)}_{\alpha,\beta} \ket{\alpha,M_A} \otimes \ket{\beta,M-M_A},
\label{eq:BD_randomstate}
\end{equation}
where the limits of the sum over $M_A$ are the same as in Eq.~\eqref{eq:factorization}, and $\psi^{(M_A)}_{\alpha,\beta}$ are uncorrelated random numbers upto normalization. 

The reduced density matrix of such a state in subsystem $A$ is of block diagonal form, $\rho_{A,M}= \sum_{M_A} p_{M_A} \rho_{A,M_A}$, where the factors $p_{M_A}\ge 0$ come from normalizing $\rho_{A,M_A}$ in each $M_A$-sector and satisfy $\sum_{M_A} p_{M_A} = 1$. The probability to find $M_A$ particles in $A$ is given by $p_{M_A}$, which is thus interpreted as the (classical) probability distribution of particle number in $A$. The entanglement entropy can then be expressed as 
\be
S(\rho_{A,M}) = \sum_{M_A} p_{M_A} S(\rho_{A,M_A})- p_{M_A} \log p_{M_A},
\label{eq:EE_sector}
\ee
where the second term on the RHS is the Shannon entropy of the number distribution $p_{M_A}$, which captures particle number correlations between the two halves, while the first term captures quantum correlations between configurations with a fixed particle number~\cite{2019Science_EEmbl}.

The uniform measure on $\mathcal{H}(M)$ was derived in \cite{2019PRD_BianchiDona} 
and is the product of the distribution on the $p_{M_A}$'s, and the uniform Haar measure within each number sector. The resulting analytical expression for the first two moments of the EE distribution for random states of the form Eq.~\eqref{eq:BD_randomstate} is reproduced in Appendix~\ref{app:exactmoments} (as a function of $L, M$), and these exact results are used in our numerical comparison below. In this section, we again discuss these moments in the limits $L_A,L_B\gg 1$, for which asymptotic forms were derived in ~\cite{Bianchi-VidmarReview}. In these limits,  the average entanglement entropy of the BD distribution is given by 
\begin{align}
   \mu_{\rm BD}(f,m)\approx &fL[ -(1-m)\log(1-m)-m\log(m)] \notag\\
    &- \sqrt{L}\delta_{f,1/2}\sqrt{\frac{m(1-m)}{2\pi}}  \bigg| \log(\frac{1-m}{m}) \bigg|\notag \\
    &+ \frac{ f+\log(1-f)}{2} - \frac{1}{2} \delta_{f,1/2} \delta_{m,1/2}, \label{eq:SMMeanscaling}
\end{align}
where  $m = M/L$. The first term is the volume law term that scales proportionately with $L_A = fL$, and the prefactor accounts for the reduced Hilbert space dimension in symmetry sector $M$. When $f=1/2$, the EE has an additional $\sqrt{L}$ contribution which comes from a saddle point evaluation of the probability distribution of $m$, and which has also been discussed in \cite{2017PRL_EEchaotic, Murthy_2019}. This correction can make finite-size analysis more complicated, but it vanishes at half-filling, $m=1/2$, which is the maximum entropy case that will be of interest to us in our comparisons with infinite temperature eigenstates. 

Evaluating Eq.~(\ref{eq:SMMeanscaling}) at half-filling ($m=1/2$) and for equal bipartitions ($f=1/2$) yields:
\begin{align}
   \mu_{\rm BD}(f=1/2,m=1/2)\approx &fL\log(2) -\frac{1}{2} + \frac{ 0.5+\log(0.5)}{2}.  
 \label{eq:BD_meancorrection}
\end{align}
Relative to the Page entropy at $f=1/2$, Eq.~(\ref{eq:Savg}), this expression has an ``extra" deficit of size $\left|\frac{ 0.5+\log(0.5)}{2}\right| \approx 0.0966$ (see relative shifts in Page and BD distributions plotted in Fig.~\ref{fig:midspectrum}; numerical values for $\mu_{\rm P}$ and $\mu_{\rm BD}$ showing the 0.1 difference for $L$=8 to 16 is shown in table I of Appendix \ref{app:exactmoments}). As we will show in the next section, this shift accounts for the $O(1)$ deviations between infinite temperature eigenstates of local Hamiltonians (without any additional symmetries) and the Page entropy that have been previously noted in the literature~\cite{2022PRE_deviationsfromETH, huang_hamentanglement, huang_midspectrum}.

The second moment of the BD distribution in the $L_A, L_B \gg 1$ limit is approximated by 
\begin{align}
    \sigma_{\rm BD}^2(f,n) &\approx \alpha L^{3/2} e^{-\beta L},
    \label{eq:SMVarScaling}
\end{align}
where $\beta=-m\log m-(1-m)\log(1-m)$ and $\alpha$ is an $O(1)$ numerical prefactor in the limits of interest. At $m=f=1/2$, the variances for the Page and BD distributions scale similarly with system size, {\it i.e.}, as $\sim 1/d^2_B \sim 2^{-L}$. Away from this limit, the rate of exponential decrease $\beta$ is different for the two distributions. However, similarly to Page, $\sigma_{\rm BD}/\mu_{\rm BD} \ll 1$ so that the average entropy of a constrained pure state in a fixed $M$ sector is also typical. 

Finally, we note that we will compare the BD distribution to the EE distribution produced by eigenstates of the MFIM model in Eq.(\ref{eq:TFIM}). The MFIM has time-reversal symmetry and thus its eigenstates are real-valued vectors, whereas the BD distribution was derived for complex random states. Thus, the reference $\mu_{\rm BD}$ and $\sigma_{\rm BD}$ need to be adjusted. For the case of unconstrained random states (i.e. without any symmetry), it has been found\cite{2010JPA_eermt,2011JPA_rmtee,2016PRE_entanglementdispersion} that the distribution of EE for both real and imaginary pure random states asymptotically has the same mean value, given by the Page mean, $\mu_{\rm P}^{\rm GOE}\approx \mu_{\rm P}^{\rm GUE}$, Eq.~\eqref{SAvgExact}. On the other hand, the standard deviation of the EE distribution, $\sigma_{\rm P}$, is (asymptotically) larger by a factor of $\sqrt{2}$ for real random states~\cite{2010JPA_eermt,2011JPA_rmtee,2016PRE_entanglementdispersion}, $\sigma_{P}^{\rm GOE} \approx \sqrt{2} \sigma_{P}^{\rm GUE}$ where GOE and GUE refer to the orthogonal and unitary ensembles applicable for real and complex random states. The exact finite-size expressions for the mean and standard deviation of the EE of real random states look significantly more complicated than the expressions for the Page distribution for complex random states, but we show in Appendix~\ref{app:exactmoments} that numerically obtained values for the mean and standard deviation of real random states converge to $\mu_{\rm P}^{\rm GUE}$ and $\sqrt{2} \sigma_{\rm P}^{\rm GUE}$ with increasing $L$. Turning to constrained states, exact analytic results for real random states with charge conservation, i.e. the GOE version of the BD distribution, have not yet been derived. However, similar to the Page case, we find numerically in Appendix~\ref{app:exactmoments} that the means of the EE distribution produced by real and imaginary states in a given symmetry sector converge to the same value with increasing $L$, while the standard deviation is again a factor of $\sqrt{2}$ larger for real states. In what follows, whenever we refer to $\sigma_{\rm BD}$, we are referring to $\sigma_{\rm BD}^{\rm GOE}$, which is inflated by a factor $\sqrt{2}$ relative to the exact expression for $\sigma_{\rm BD}$ in Eq.~\eqref{VarSAFixedM}, while we continue to use the expression for $\mu_{\rm BD}$ in Eq.~\eqref{MeanSAFixedM} for the mean\footnote{We have checked that our results are not qualitatively changed on using these analytic reference values, as compared to using numerically determined moments for real random constrained states.}.

\subsubsection{Application of the BD distribution to midspectrum Hamiltonian eigenstates}

We have seen in the previous section that the approach to compare the eigenstates of FRCs with unconstrained pure random states captures certain properties like the average entropy, while deviations from RMT due to locality are dominantly reflected in higher moments. 

To generalize this analysis to Hamiltonian systems, we must contend with the fact that energy conservation adds additional structure which is not captured by the Page distribution, and which already results in finite deviations in the average half-system entropy of infinite temperature eigenstates~\cite{2022PRE_deviationsfromETH,huang_hamentanglement, huang_midspectrum, huang_deviation_2022} (see Fig.~\ref{fig:midspectrum}). The goal is to identify a new (more constrained) random state distribution which incorporates the effect of energy conservation, so that differences between the new reference distribution and the microcanonical distribution can be dominantly attributed to features such as locality.

We now argue for why the BD distribution furnishes a better RMT ensemble for describing midspectrum eigenstates of local Hamiltonians, even in the absence of additional $U(1)$ symmetries like particle number. The effect energy conservation alone can be captured, within certain approximations, by that of conservation of a local additive $U(1)$ charge, and the BD distribution incorporates this feature. Our arguments recapitulate and build on part of the discussion in Ref.~\cite{huang_midspectrum}, which argued for an $O(1)$ deviation between the Page entropy and mean EE of Hamiltonian eigenstates.   

As mentioned, we will be interested in the eigenstate entanglement entropy $S_A$ of a subsystem of size $A$. We write the Hamiltonian as $H = H_A+H_B+H_{AB}$, where $H_A$ ($H_B$) has support on $A$ ($B$) only, and $H_{AB}$ has support in both $A$ and $B$. We can write any eigenstate of $H$ with energy $E$, $\ket{\psi_E}$, in the basis of tensor products of eigenstates of $H_A, H_B$:
\begin{equation}
\ket{\psi_E} = \sum_{ij} c_{ij} |\epsilon_i\rangle_{\tiny A}  |\epsilon_j\rangle_{\tiny B}  
\label{eq:deutsch}
\end{equation}
Deutsch proposed~\cite{1991PRL_Deutsch} that Hamiltonian eigenstates could be modeled as random states in which $c_{ij}$ is a random matrix with a narrow bandwidth which approximately imposes that the sum of energies of the subsystems is approximately equal to $E$:  $\delta E_{ij} \equiv  \epsilon_i + \epsilon_j - E \approx 0$ (the equality is not exact because of $H_{AB}$). Ref.~\cite{Murthy_2019} used ETH to refine this conjecture to a more explicit form in which $c_{ij}$ is modulated by a ``window function" $F(\delta E_{ij})$ which penalizes deviations away from $\delta E=0$ on a scale set by $\Delta = \sqrt{\langle \psi_E|H_{AB}|\psi_E\rangle }$.  

With the condition $\delta E_{ij} \approx 0$,  Eqs~\eqref{eq:BD_randomstate} and \eqref{eq:deutsch} are conceptually 
very similar. However, there are also differences that we must treat with caution: (i) The presence of the $H_{AB}$ term (which is also responsible for the state being entangled in the first place) means that the $\delta E_{ij}$ is only approximately (rather than exactly) equal to 0. (ii) the spectrum of $H$ is dense, so that the Hilbert space does not factor into a sum of tensor products as in Eq.~\eqref{eq:EE_sector}. 

Both these differences can be addressed if we make an approximation which sets the ``window function" $F(\delta E_{ij})$ to be strictly zero outside some width $\delta$ set by $\Delta$, so that $c_{ij}$ is a strictly banded random matrix. This is well-motivated, also because of a mathematical proof in Ref.~\cite{arad2016connecting} which shows tha,t for local Hamiltonians, there exist constants $c,\delta >0$ such that $\sum_{\Delta E_{\ij} \geq \Lambda} |c_{ij}|^2 < c e^{-\Lambda/\delta}$. The truncated state  can now be put in the form of Eq.~\eqref{eq:BD_randomstate} (note that $\psi^{M_A}_{\alpha, \beta}$ in \eqref{eq:BD_randomstate} is a banded random matrix). This is because the \emph{same} truncated state can be equivalently obtained by discretizing the spectrum of $H_A, H_B$ in steps of size $\delta$, so that all eigenvalues with energy $ E_i- \delta/2 \leq  E < E_i+\delta/2$ are assigned to the $i$th step $E_i$. This produces a degeneracy for $E_i$, similar to the degeneracy in $M_{A/B}$ for the $U(1)$ case.
Then, the truncated state is just a random constrained state with the \emph{strict} constraint that $E_i +E_j=E$, which now looks identical to the $U(1)$ constrained state of the previous section in which the step size was $1$. This factors the Hilbert space into a sum of tensor products, as in Eq.~\eqref{eq:EE_sector}.  
The relative dimensions of the steps at different energies also become equivalent to the $U(1)$ case in the large system limit, since  the density of states of the Hamiltonian approaches a Gaussian, as does the binomial ``choose" function which sets the sizes of the $U(1)$ sectors. 

We emphasize that the locality of the Hamiltonian is crucial for making the connection between the eigenstate and BD distributions. The truncation scale $\Delta$ is set by $H_{AB}$, and we need  $\Delta \sim O(1)$ for the arguments above. More colloquially, we want $H_{AB}$ to be a weak boundary term which is necessary to couple the subsystems, but can be taken to be arbitrarily small while still getting a thermal state.  
In contrast, we expect that eigenstates of long-range  {or k-local} models ({\it i.e.}, SYK models or systems {with} power law interactions) will have the same universal properties as pure random states without any constraints, and  defer a more detailed analysis of this to future work.

Finally, we note that our arguments above do not rule out the possibility of non-universal O(1) corrections in the mean EE induced by the truncation of the $c_{ij}$ matrix and the presence of the $H_{AB}$ term. Nevertheless, we find below that the agreement between BD and the Hamiltonian eigenstate distribution is surprisingly good, and captures most of the observed difference between the eigenstate EE and Page mean. For this reason, we conjecture that the BD distribution is the best {\it universal} distribution to incorporate the effects of energy conservation.

\subsection{The Mixed Field Ising model (MFIM)}
\label{sec:mfim}

We now describe the Hamiltonian model studied in this work, the one-dimensional MFIM:
\be
H = \sum_{i} \left(\sigma_i^z\sigma_{i+1}^z + g\sigma_i^x + h \sigma_i^z\right),
\label{eq:TFIM}
\ee
where $\sigma_i^\alpha$ ($\alpha = x,y,z$) are Pauli matrices, $g$ is the transverse field, and $h$ is the longitudinal field. We use open boundary conditions in order to break translational symmetry, and add additional boundary fields $h_1 = 0.25$ and $h_L = -0.25$ at the edges to break inversion symmetry.\footnote{The results are not sensitive to the value of symmetry breaking field.} 
The MFIM has various limits of physical interest. When $h=0$, the model can be mapped to a free fermion model through a Jordan-Wigner transformation~\cite{1964RMP_jordanwigner}, therefore the model is non-interacting and integrable. A finite value of $h$ breaks integrability. In addition, the model hosts two classical integrable limits: (i) $g=0$ corresponds to the classical Ising model (diagonal in the $\sigma^z$ basis), and (ii) $g\gg 1$ corresponds to the classical paramagnet (diagonal in the $\sigma^x$ basis).  

The MFIM has been extensively studied numerically in the context of thermalization and chaos~\cite{Zhang_2015,Roberts_2015, Kim_2013}, with diagnostics ranging from eigenstate entanglement entropy to level spacing ratio~\cite{Zhang_2015} to entanglement growth~\cite{Kim_2013} and operator spreading dynamics~\cite{Roberts_2015,KhemaniU1}. These numerical studies have largely worked with two parameter choices that have been identified as showing particularly strong thermalizing behavior even at relatively small sizes: The `Banuls-Cirac-Hastings (BCH)' parameters~\cite{BanulsCiracHastings}, $g= -1.05$ and $h = 0.5$, and the `Kim-Huse (KH)' parameters~\cite{Kim_2013}, $g = (\sqrt{5}+5)/8 \approx 0.9045$ and $h= (\sqrt{5}+1)/4 \approx 0.8090$. These choices have become standard in the literature, and we will refer back to them once we discuss our numerical results. In particular, we find that these points are \emph{not} the most chaotic with respect to the more resolved metric of chaos we present, even as various other standard diagnostics  look chaotic at these parameter values. 

In what follows, we focus on the distribution of entanglement entropy of midspectrum eigenstates for a half subsystem, $f=1/2$, centered in the middle of the system.  We use exact diagonalization to obtain the entire spectrum for system sizes up to $L=14$, and fit the density of states (DOS) to find the energy corresponding to the peak of the DOS, {\it i.e.}, to infinite temperature (since $\Tr(H)=0$, the value of energy corresponding to infinite temperature approaches zero with increasing size, but finite size systems can show small deviations within the scale of energy fluctuations). We then obtain the EE of all eigenstates centered in a small energy window around the peak energy, and compute the mean and standard deviation of the EE of these eigenstates, denoted $\mu_{ H}$ and $\sigma_{ H}$ respectively. Here the subscript $H$ refers to `Hamiltonian', to be contrasted with Floquet unitaries studied in the previous section, and is thus more descriptive than the $\mu_{\rm E}$, $\sigma_{\rm E}$ symbols introduced in Sec.~\ref{sec:metric} which referred collectively to eigenstate distributions of either Hamiltonian or Floquet systems.

The number of eigenstates within the energy window centered at the DOS peak is chosen large enough to minimize the uncertainty of $\mu_H$ and $\sigma_H$, but small enough to avoid systematic effects induced by finite temperature eigenstates. Here we use windows with 100, 400, and 600 states for $L=10$, $L=12$, and $L=14$, respectively. For larger system sizes ($L=16$), we do not diagonalize the full spectrum but, instead, use the shift-invert method~\cite{2018SciPost_shiftinvert} to find 2000 eigenstates closest to $E=0$. Our results are quite robust to changes in the number of states chosen, and a discussion on selecting a window size with an appropriate number of states is presented in Appendix~\ref{app:windowsize}. 

For comparison, we also show data for the level spacing ratio~\cite{oganesyan_localization_2006} averaged over the same energy windows defined above. The ratio factor is a commonly used diagnostic of level repulsion and is defined using three consecutive eigenstates, $\{E_{n-1}, E_n, E_{n+1}\}$, as: 
\be
0 \le r_n = \frac{{\rm min}(\Delta E_n,\Delta E_{n+1})}{{\rm max}(\Delta E_n,\Delta E_{n+1})}\le 1,
\ee
with $\Delta E_n = E_n-E_{n-1}$. Integrable systems exhibit uncorrelated level statistics described by a Poisson distribution with average ratio factor $\langle r \rangle = 0.386$ \cite{atas_distribution_2013}. Chaotic systems exhibit level repulsion described by Wigner-Dyson statistics with an average ratio factor  $\langle r \rangle = 0.536$ \cite{atas_distribution_2013}.

\begin{figure}
\centering\includegraphics[scale = 1.0]{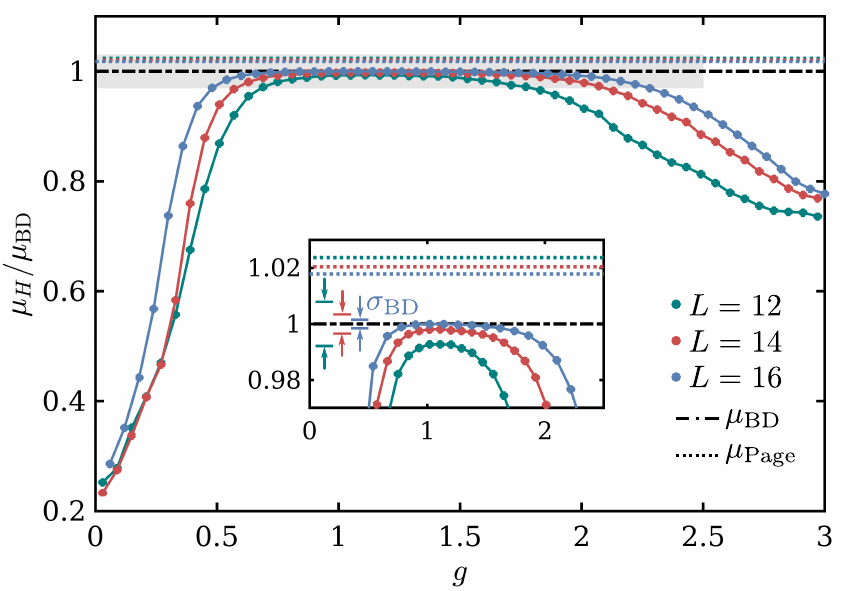}
\caption{
Finite-size scaling of the first moment $\mu_H$ of the microcanonical entanglement entropy distribution of Hamiltonian eigenstates plotted as a function of the transverse field $g$ in the MFIM. Results are normalized with respect to the first moment $\mu_{\rm BD}$ of the BD distribution. Curves are plotted for system sizes $L = 12,14,16$ and a longitudinal field $h=0.3$. We see that the microcanonical mean converges to $\mu_{\rm BD}$. The inset shows a zoomed version of the main panel (see shaded region) showing approach to the BD distribution (dotted dashed lines) and statistically significant deviations from the Page distribution (dotted lines). Indicated with arrows is the standard deviation $\sigma_{\rm BD}$ of the BD distribution for the different system sizes.}
\label{fig:meanvsstd}
\end{figure}

\subsection{First moment of the EE distribution}

We begin our discussion by focusing only on the first moment of the microcanonical EE distribution $\mu_H$ and study its system size dependence. Figure~\ref{fig:meanvsstd} shows $\mu_{ H}$ as a function of the transverse field $g$ while keeping $h$ fixed to $h=0.3$. The choice of $h=0.3$ corresponds to a strongly chaotic cut, as will become clear in the next subsection.  All curves are normalized by the theoretical value of the first moment of the BD distribution, $\mu_{\rm BD}$, obtained from Eq.~\eqref{MeanSAFixedM} of Appendix~\ref{app:exactmoments} and evaluated at half-filling (maximum entropy), $m=1/2$,  and equal bipartition, $f=1/2$.

We find excellent agreement between the means $\mu_H$ and $\mu_{\rm BD}$ in a wide range of parameters centered around $g\approx 1.1$. In particular, for $L=16$ we find agreement between $\mu_H$ and $\mu_{\rm BD}$ up to the fifth significant digit. In addition, by increasing the system size,  we observe that $\mu_H$ approaches $\mu_{\rm BD}$ in an increasingly larger region of parameter space: although there are sizable deviations between $\mu_H$ and $\mu_{\rm BD}$ 
close to integrabile limits ($g\ll 1$ or $g\gg 1$), these tend to decrease with increasing system size. To leading order, this behavior agrees with that observed using spectral metrics of chaos: the MFIM exhibits chaotic behavior for all finite values of $g$ in the thermodynamic limit. 

In contrast, we observe statistically significant deviations between $\mu_H$ and the mean EE of pure random states without any constraint $\mu_{\rm P}$ (dotted lines) for {\it all} values in parameter space and for all system sizes, consistent with recent observations~\cite{2022PRE_deviationsfromETH,huang_hamentanglement, huang_midspectrum}. Such behavior
is visible in the inset of Fig.~\ref{fig:meanvsstd}, which shows a zoomed version of the main panel with  $\mu_{\rm P}$, Eq.~\eqref{SAvgExact}, plotted with dotted lines. These results provide strong numerical corroboration that, in contrast to the Page distribution, the BD distribution is a better reference RMT distribution for midspectrum eigenstates of local Hamiltonians.  

At the level of the first moments, the Hamiltonian results are analogous to the FRC results: pure random states (with appropriate constraints) correctly describe the first moment of the EE distribution of eigenstates in quantum chaotic systems. In the following two subsections, we use the more refined metric $D_{\rm KL}$ that compares differences on the exponentially small scale set by $\sigma_{\rm R}$ and also incorporates the effects of the second moment of the EE distributions.

\begin{figure}
\centering\includegraphics[scale = 1.0]{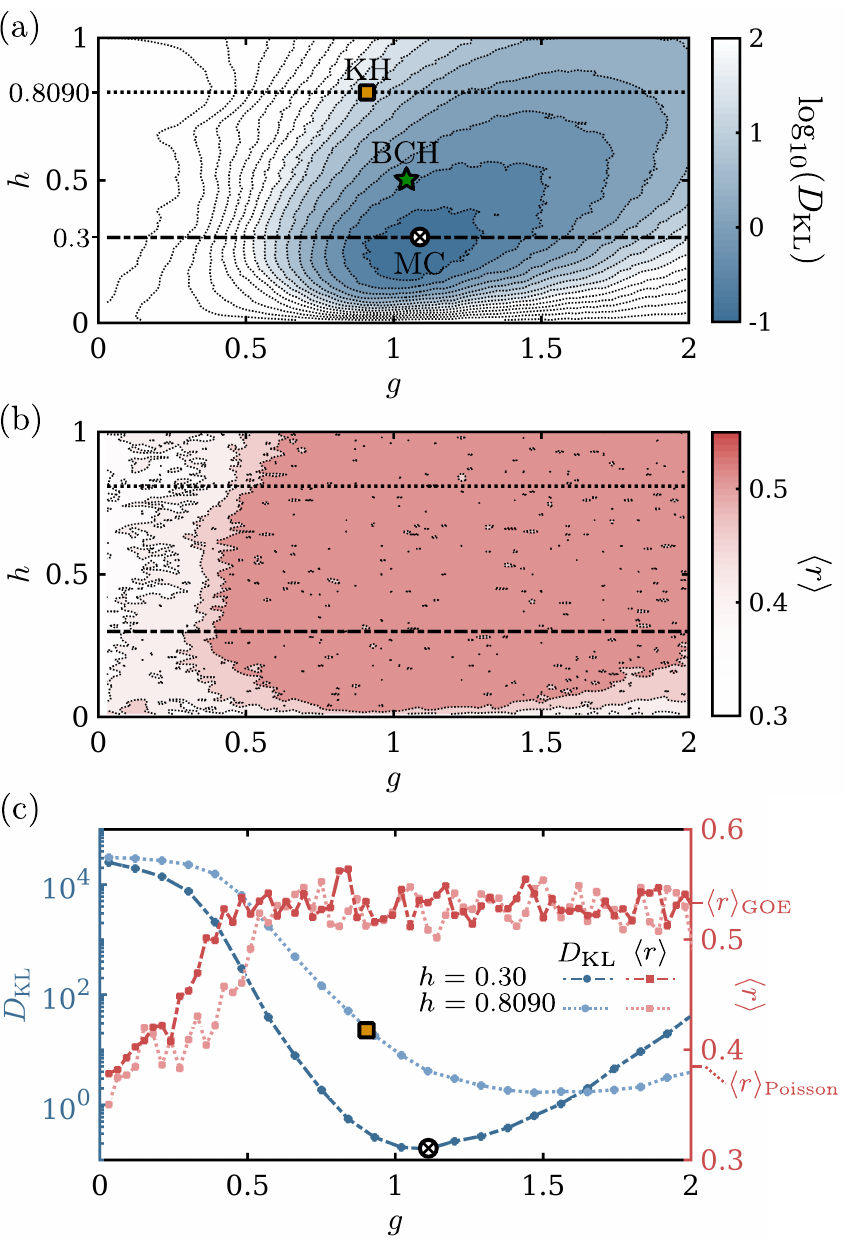}
\caption{Colormaps of (a) the KL divergence and (b) the average ratio factor $\langle r \rangle$ computed for the MFIM with transverse field $g$ and longitudinal field $h$, for $L=14$. The $D_{\rm KL}$ contour plot shows that, within the parameter space of the MFIM model, the parameter values near $(g_*,h_*)=(1.1\pm 0.05,0.30\pm 0.05)$ are most chaotic, with $D_{\rm KL}$ steeply increasing away from these values. Also indicated are the Kim-Huse (KH, square) and Banuls-Cirac-Hastings (BCH, star) parameters that have been widely used in studies of thermalization, which show much larger values of $D_{\rm KL}$. In contrast, panel (b) shows the value of $\langle r \rangle$ is saturated at the RMT value and signals chaos in a broad region of parameter space (including the BCH and KH parameters) and does not exhibit the resolution observed in panel (a).  (c) Horizontal linecuts of $D_{\rm KL}$ and $\langle r \rangle$ across the MC and KH parameters, see horizontal dotted-dashed and dotted lines, respectively, in panels (a) and (b). The reference GOE and Poisson ratio factors are indicated on the right vertical axis.
}
\label{fig:DKL}
\end{figure}

\subsection{Kullback-Liebler divergence and maximally chaotic Hamiltonians}

Having discussed the behavior of the first moment of the microcanonical EE distribution relative to the BD distribution, we now employ our more refined measure of `distance between distributions' using the KL divergence. As discussed in Sec.~\ref{sec:metric}, we only use the first two moments to compute the KL divergence via Eq.~\eqref{eq:DKL_approx}. 

Figure\,\ref{fig:DKL}(a) shows the value of ${D}_{\rm KL}$ as a function of model parameters $(g,h)$ for $L=14$. A noticeable feature of Fig.\ref{fig:DKL}(a) is that, for most of parameter space, the eigenstate and RMT distributions exhibit relatively large deviations from each other. Indeed, there is only a small region of parameter space where the value of ${D}_{\rm KL}$ is small ({\it i.e.}, ${D}_{\rm KL} \lesssim 1$). The minimum value for ${D}_{\rm KL}$ is obtained for $(g_*,h_*) = (1.10\pm0.05,0.30\pm 0.05)$, which thus corresponds to the most chaotic (MC) parameters for this metric and this family of MFIM Hamiltonians. The KL divergence increases exponentially as $(g,h)$ are tuned away from the MC parameters, Fig.~\ref{fig:DKL}(c). The MC parameters that we find are relatively close to the BCH parameters in Ref.~\cite{BanulsCiracHastings}, but far from the KH parameters in Ref.~\cite{Kim_2013}, even though both sets of parameter choices are widely employed in studies of chaos in the MFIM. Both parameter choices are indicated in Fig~\ref{fig:DKL}(a).
In the next subsection and in Appendix~\ref{app:MFIM_litcompare}, we show a more detailed comparison between the MC parameters identified in this work, and those commonly used in the literature. 

To contrast the behavior of $D_{\rm KL}$ with other metrics of chaos, in Fig.\ref{fig:DKL}(b), we plot the average level spacing ratio $\langle r \rangle$ as a function of $g$ and $h$. We find that $\langle r \rangle$ is featureless and saturates to the RMT value  $\langle r\rangle \approx 0.54$ in a very broad region of parameter space, unlike ${D}_{\rm KL}$ which is only minimized in a small region of parameter space around $(g_*,h_*) = (1.1,0.35)$.  Fig.~\ref{fig:DKL}(c) shows two linecuts as a function of $g$, at the $h$ value corresponding to the MC and KH points. We see that $\langle r\rangle$ quickly saturates to the RMT value away from the $g=0$, while $D_{\rm KL}$ shows strong variation in parameter space even for parameters for which $\langle r \rangle $ looks strongly chaotic. This is consistent with the general picture discussed above which argues that $D_{\rm KL}$ is a much more resolved metric of quantum chaos.

The enhanced proximity to RMT behavior in a small pocket of parameter space is quite striking: this seems to indicate that `maximally' chaotic local Hamiltonians --- those with microcanonical distributions of EE reproducing the first and  higher moments of pure random state distributions ---  are not typical. Although our two-parameter model exhibits agreement with RMT in a small region of parameter space, a tantalizing possibility is that these small regions shrink to a fine-tuned point when extending the space of local Hamiltonian models to include a larger number of parameters. 
This is reminiscent of integrable models which are assumed to be fined tuned points in parameter space. It is also reminiscent of special classes of Floquet models, namely, dual-unitary Floquet models, in which the spectral form factor shows a Thouless time equal to 1 and which are therefore considered `maximally' chaotic from the lens of spectral statistics~\cite{Bertini_2018}.  While our analysis was restricted to the two-dimensional parameter space of the MFIM, an interesting question for future study is whether there exists a  Hamiltonian with more parameters (but still local) where the distance to RMT behavior is provably minimal.

\begin{figure}
\centering\includegraphics[scale = 1.0]{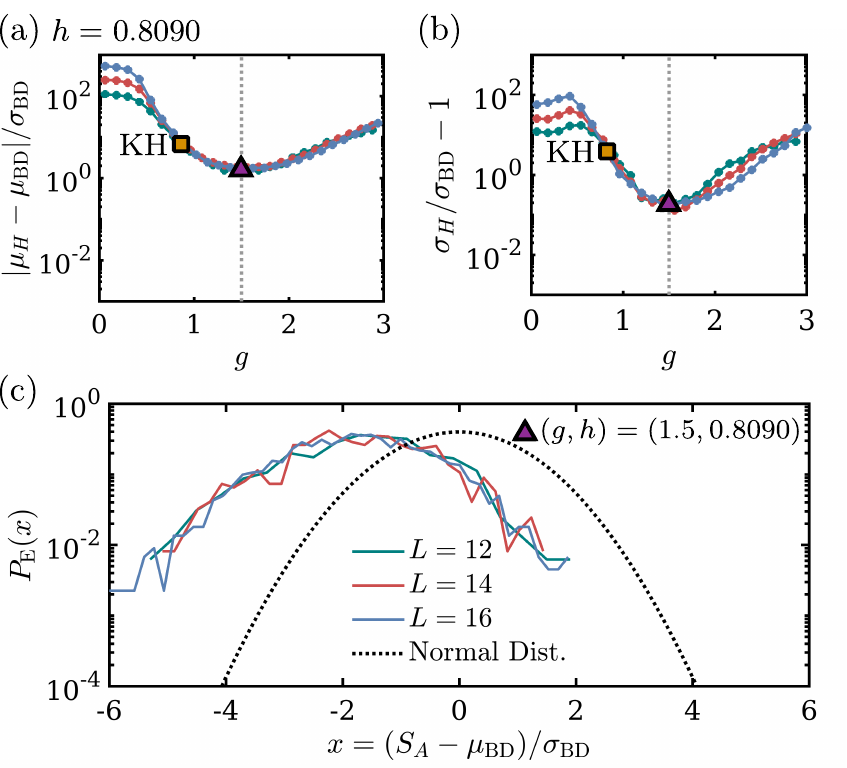}
\caption{(a) Difference between the EE mean of midspectrum eigenstates and $\mu_{\rm BD}$ normalized with the EE standard deviation of pure random states, and (b) ratio between EE standard deviation of midspecturm eigenstates and and $\sigma_{\rm BD}$  as a function of system size and transverse field $g$. Plots are shown for $L=12,14,16$, and for $h = (1+\sqrt{5})/4 \approx 0.8090$. The number of states used to compute the first two moments is discussed in the main text. The vertical dashed indicates the most chaotic point (minimal $D_{\rm KL}$) along this linecut ($g \approx 1.5$), while the square symbol indicates the KH parameters. 
(c) Histogram of midspectrum entanglement entropy for the most chaotic point along this linecut,  $(g,h) = (1.5,0.8090...)$, as a function of system size. Shown with dotted lines is the reference standard normal distribution.}
\label{fig:KH}
\end{figure}

\subsection{Higher moments of the EE distribution}
We finish our discussion of Hamiltonian systems by showing the full EE distribution of midspectrum eigenstates for various representative parameter values, and parsing how different moments of the microcanonical EE distribution contribute to $D_{\rm KL}$ for different linecuts. 

Figure~\ref{fig:KH} shows (a)  $|\mu_H-\mu_{\rm BD}|/\sigma_{\rm BD}$, and (b) $\sigma_H/\sigma_{\rm BD} -1 $ for $h = \frac{1+\sqrt{5}}{4} \approx 0.8090$ (the Kim-Huse choice) and as a function of $g$. Both ratios are normalized with $\sigma_{\rm BD}$, as they appear in the definition of $D_{\rm KL}$.  We observe substantial variation in these ratios as a function of $g$, with both quantities (and hence $D_{\rm KL}$) changing by orders of magnitude as $g$ is tuned. The moments are closest to the BD predictions near $g \approx 1.5$. Interestingly, we observe a good collapse of both ratios as a function of $L$ in the parameter region $1 \lesssim g \lesssim 2$ centered around $g\approx 1.5$.  In other words, in this regime, both the microcanonical and RMT standard deviations scale similarly with system size, $\sigma_H \sim \sigma_{\rm BD} \sim \sqrt{2^{-L}}$ but, similar to the Floquet case, $\sigma_H$ still has a systematically larger prefactor by about 20\%. Likewise, the difference between the means is exponentially small in $L$, again with the same scaling as $\sigma_{\rm BD}$. In this chaotic (but not maximally chaotic) regime, the value of $D_{\rm KL}$ appears converged with $L$ to a fixed value. 

On tuning $g$ away from the chaotic region near $g \approx 1.5$, we observe a crossover in the functional dependence of $\sigma_H$ with $L$, which goes from displaying chaotic scaling (in particular, $\sigma_H$ decreasing exponentially with increasing system size according to Eq.~\eqref{eq:SMVarScaling}) to near-integrable behavior ($\sigma_H$ decreasing polynomially with increasing system size). Accordingly, the ratio $\sigma_{H}/\sigma_{\rm BD}$ increases exponentially with $L$ near-integrability ($g$ small). Similarly, the ratio $|\mu_H - \mu_{\rm BD}|/\sigma_{\rm BD}$ increases exponentially with $L$ in the same near-integrable regime, where the difference in microcanonical and BD means is only polynomially converged.  Remarkably, we see that the sensitivity of the two ratios observes deviations from RMT over a large range of $g$'s, with the ``near-integrable" scaling observed throughout the window $0 \leq g\lesssim 0.8$, and persisting all the way till $g\sim 0.8$, which is comparable to $J$. These more-sensitive metrics also reveal that the Kim-Huse parameter choice $g\approx 0.9$ (marked by a square in Figure~\ref{fig:KH}) is surprisingly near the crossover from the integrable to chaotic scaling in the ratios, in contrast to measures like $\langle r\rangle$ which look strongly chaotic for the KH parameters. 

We note that previous works have employed the crossover from power-law to exponential scaling of the fluctuations of local operators as a way to distinguish chaotic from integrable behavior~\cite{2020PRX_Sels_chaos}. This was shown to be a more sensitive probe of chaos as compared to spectral metrics such as the ratio factor, in the sense of detecting chaos \textit{before} the ratio factor when tuning away from an integrable point.  In contrast, we use our ratios to detect deviations from chaos even in regimes where  $\langle r\rangle$ looks strongly thermal. 

In Fig.~\ref{fig:KH}(c), we show the full distribution of EE of midspectrum eigenstates, appropriately shifted and rescaled, for $g=1.5$, which is the most chaotic point with minimal $D_{\rm KL}$ along this linecut (denoted by a triangle in Fig.~\ref{fig:KH}). The rescaled distributions appear to be well converged with system size. Compared to the FRC data in Fig.~\ref{fig:FRC}, we notice that the mean \emph{also} shows a sizable departure, and is several standard deviations away from $\sigma_{\rm BD}$. 

Indeed, in most of the parameter space away from maximally chaotic (MC) point marked in Fig.~\ref{fig:DKL}, we notice that the main contribution to $D_{\rm KL}$ comes from the first moment, although $D_{\rm KL}^{(1)}$ and $D_{\rm KL}^{(2)}$ closely track each other in their qualitative behavior as a function of $g$. Given this observed departure in the first moment, in Appendix~\ref{app:sigmaratio} we also present a related diagnostic of chaos which is agnostic to the reference RMT distribution, and only looks at the (normalized) fluctuations of EE: $\sigma_H/\sqrt{2^{-L}}$. This ratio is expected to be system-size independent and minimized for maximally chaotic systems, while being exponentially increasing for $L$ for near integrable systems. Fig.~\ref{fig:sigmaratio} in Appendix~\ref{app:sigmaratio} shows that this ratio yields a qualitatively similar landscape of chaos in parameter space as $D_{\rm KL}$ in  Fig.~\ref{fig:DKL}(a). 

\begin{figure}
\centering\includegraphics[scale = 1.0]{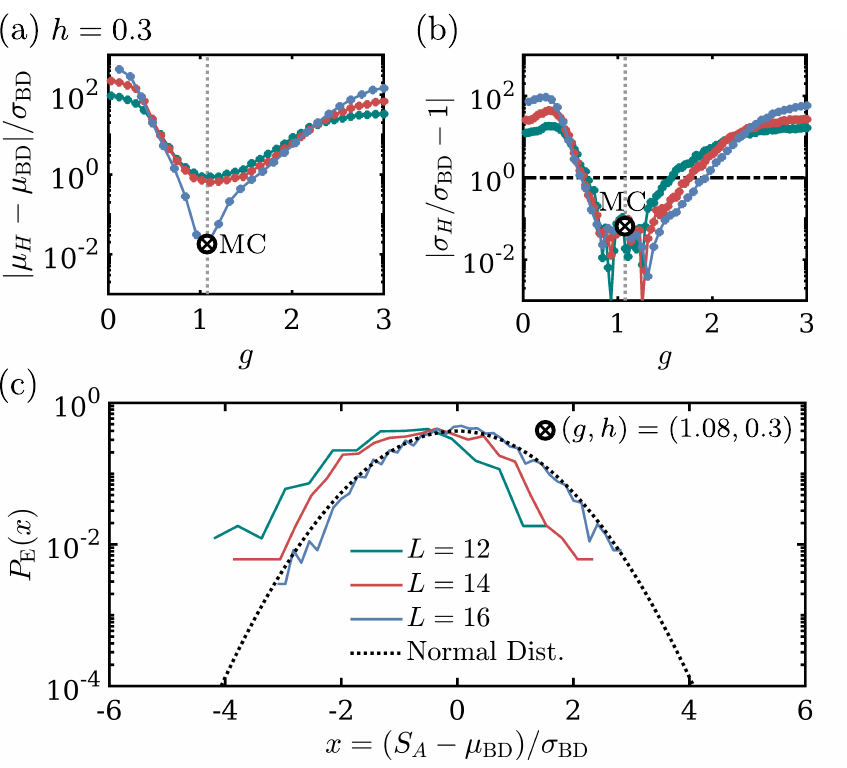}
\caption{(a) Difference between the EE mean of midspectrum eigenstates and $\mu_{\rm BD}$ 
normalized by $\sigma_{\rm BD}$, and (b) ratio between EE standard deviations of midspectrum eigenstates and $\sigma_{\rm BD}$ as a function of system size and transverse field $g$. Plots are shown for $L=12,14,16,$ and for $h = 0.3$. The number of states used to compute the first two moments is discussed in the main text. The vertical dashed indicates the most chaotic point (minimal $D_{\rm KL}$) along this linecut ($g =1.08$), which also coincides with the most chaotic (MC) parameters in the 2D parameter space.  (c) Histogram of midspectrum EE for the most chaotic (MC) parameters as a function of system size, showing approach to RMT behavior.  Shown with dotted lines is the reference standard normal distribution.}
\label{fig:MC}
\end{figure}

Next, in Fig.~\ref{fig:MC} we plot the same ratios as in Fig.~\ref{fig:KH}, but for a cut at $h=0.3$ which includes the maximally chaotic point which minimizes $D_{\rm KL}$ in the $(g,h)$ parameter space. Again, in the near-integrable regimes ($g \approx 0$ and $g \gg h$), the ratios involving both moments show an exponentially increasing trend with $L$, for identical reasons to those discussed above. There is, however, a notable difference from Fig.~\ref{fig:KH} in the parameter regime near the MC point at $g \approx 1.1$. In particular, we find that both ratios show a sharp change for the largest size ($L=16$) to show much better convergence with the BD predictions at the MC point, and the ratio involving the first moment shows a steep increase away from the MC point as shown in Fig.~\ref{fig:MC}(a). In fact, we also observe that for a small range of parameters near $g = 1.1$, the microcanonical standard deviation is even slightly smaller than the reference $\sigma_{\rm BD}$, which is why we plot the absolute value of the deviation of this ratio from 1 in Fig.~\ref{fig:MC}(b). The full distribution of midspectrum eigenstates for the MC parameters is shown in panel (c), showing almost perfect agreement with the BD distribution at $L=16$, even more so than the FRC data Fig.~\ref{fig:FRC} which showed an observable difference in the second moment. 

This near-perfect agreement with the BD distribution at the MC point and the sharp change in behavior with $L$ and $g$ near this point is quite surprising --- {\it a priori}, one might have expected deviations from BD everywhere in parameter space given that the Hamiltonian is local and only has two tuning parameters. We might also have expected a gentler change in behavior in parameter space away from the MC point, contrary to what is observed in Fig.~\ref{fig:MC}(a). We defer a more detailed analysis of the MC point and perturbations away from it to future work.

\section{Discussion and outlook}
\label{sec:discussion}

We introduce a quantitative metric of chaos which measures the Kullback-Liebler between the microcanonical distribution of EE generated by eigenstates of local Hamiltonian/Floquet systems, and a reference random distribution. This metric serves as a much more resolved measure of chaos, which compares both the mean and the standard deviations of the microcanonical and reference distributions on the exponentially small scale set by $\sigma_{\rm R}$. This not only distinguishes between integrable and chaotic behavior, but also furnishes a continuous ``ruler" which measures deviations from RMT even as other spectral metrics such as the level spacing ratio look strongly thermal. 

Besides introducing a new method for characterizing thermalization, we emphasize several other ramifications of our work. 
First, we show that the distribution of entanglement entropy deviates from RMT predictions even in paradigmatic models of strongly thermalizing dynamics, namely, Floquet Random Circuits (FRC) without any structure other than locality. This is primarily reflected in the ratio $\sigma_U/\sigma_{\rm P} > 1$, which is larger than 1 and stable with increasing system size.  An interesting direction for future work is to understand how different physical effects stemming from locality contribute to the increased standard deviation. For example, prior works~\cite{Dymarsky_deviation, 2022PRL_dymarsky, 2020PRE_beyondETH,2021PRE_eth_otocs, 2019PRE_kurchan,2019PRL_Chalker, Garratt_2021}  have shown that the existence of a light-cone in the spreading of operators in local quantum systems leads to beyond-RMT spectral correlations in the eigenstate expectation values of local operators. It would be fruitful to try to establish a quantitative connection between these correlations and the increased standard deviation of the microcanonical ensemble. Locality also implies that the entangling properties of the local gates in the FRC (those that straddle the entanglement cut) plays an important role in the eigenstate entanglement.

Second, our work provides evidence that a more suitable reference distribution for comparing eigenstates of local Hamiltonian systems is the Bianchi-Dona distribution (which accounts for the presence of a U(1) charge) as opposed to the Page distribution, which has been the standard reference distribution in almost all previous works. In particular, building on Ref.~\cite{huang_midspectrum}, we argue that local Hamiltonians with energy conservation effectively have a local scalar charge which behaves similarly to the U(1) charge for infinite temperature eigenstates and for large enough subsystems. An important direction for future work is to better understand the effects of finite temperature, locality, and symmetries (including multiple, possibly non-commuting symmetries) in setting the appropriate reference RMT ensemble for Hamiltonian/Floquet eigenstates, and to try to incorporate the universal contributions of these features into the reference distribution.  
Separate from the challenge of finding an appropriate reference distribution, we showed that the microcanonical fluctuations of EE (normalized by $\sqrt{2^{-L}}$ serves as an independent (and reference agnostic) diagnostic of chaos which qualitatively displays similar behavior to $D_{\rm KL}$ in parameter space and is minimized for the most chaotic models.

Finally, our results show a strong and surprising variation of $D_{\rm KL}$ in parameter space for the MFIM, showing strong deviations away from RMT even for parameters that had been previously identified as strongly chaotic, and in parameter regimes where other metrics have saturated to chaotic values. Conversely, we observe that $D_{\rm KL}$ is minimized in small pockets of parameter space, suggesting that there might be families of ``maximally chaotic" Hamiltonians. This is somewhat reminiscent of ``minimally choatic" (integrable) systems which are known to be fine-tuned points of measure zero in parameter space, or maximally chaotic dual-unitary Floquet circuits. More detailed studies about these maximally chaotic regions (or points), and understanding their dynamical properties remains an open direction for future research.

\section*{Acknowledgements}

We are grateful to John Chalker, Wen Wei Ho, Nick Hunter-Jones, David Huse, Matteo Ippoliti and Chaitanya Murthy for insightful discussions. JFRN acknowledges the Gordon and Betty Moore Foundation’s EPiQS Initiative through Grant GBMF4302 and GBMF8686 for a postdoctoral fellowship at Stanford University. This work was supported by the US Department of Energy, Office of Science, Basic Energy Sciences, under Early Career Award Nos. DE-SC0021111 (C.J. and V.K.).  V.K. also acknowledges support from the Alfred P. Sloan Foundation through a Sloan Research Fellowship and the Packard Foundation through a Packard Fellowship in Science and Engineering. Numerical simulations were performed on Stanford Research Computing Center's Sherlock cluster.  We acknowledge the hospitality of the Kavli Institute for Theoretical Physics at the University of California, Santa Barbara (supported by NSF Grant PHY-1748958).


%

\appendix

\renewcommand{\thefigure}{A\arabic{figure}}
\renewcommand{\theequation}{A\arabic{equation}}
\setcounter{equation}{0}
\setcounter{figure}{0}


\section{Moments of the entanglement entropy distribution for pure random states}
\label{app:exactmoments}

\subsection{The Page distribution for pure random states}

The distribution of entanglement entropy averaged over pure random states was computed analytically in several works~\cite{2019PRD_BianchiDona, 2016PRE_entanglementdispersion,2017PRE_entanglementvariance_proof}. A pure state $\ket{\psi}$ on the composite system $\mathcal{H} = \mathcal{H}_A\otimes \mathcal{H}_B$ can be expanded as a linear combination
\begin{align}
    \ket{\psi}=\sum_{i=1}^{d_A} \sum_{j=1}^{d_B} C_{i,j} \ket{i}\otimes \ket{j},
\end{align}
where $d_A$ ($d_B$) is the Hilbert space dimension of system $A$ ($B$), and the coefficients $C_{i,j}$ are the entries of a rectangular $d_A \times d_B$ matrix $C$. The reduced density matrix for such a state after tracing out the degrees of freedom in system $B$ is
\begin{align}
    \rho_A&= \sum_{i,i'=1}^{d_A} W_{ii'} \ket{i}\bra{i'} \label{rhoA}, 
\end{align}
with $W_{ii'} = \sum_j C_{ij}C_{i'j}^* = (CC^\dagger)_{ii'}$. The entanglement entropy $S(\rho_A)=\sum_i \lambda_i \ln \lambda_i$ for $|\Psi\rangle$ is determined from the spectrum $\{\lambda_i\}$ of $W$. For random states the coefficients $C_{i,j}$ are independently and identically distributed real (GOE) or complex (GUE) Gaussian variables, following the distributions $P(C)\propto$ exp$\{-\frac{\beta}{2}$ Tr$(C^{\dagger}C)\}$, where $\beta = 1$ for the GOE ensemble and $\beta = 1$ for the GUE ensemble. In such cases, the product $W=CC^{\dagger}$ is known as a random Wishart matrix, and the joint probability distribution $P(\{\lambda_i\})$ has been well characterized~\cite{2004Mehta_RMT}. This allows one to compute all the moments of $S_A$ for real and random state ensembles. 

The statistical properties of \textit{complex} random states, which are descriptive of eigenstates of FRC or systems with broken time-reversal symmetry (TRS), are better understood than those of \textit{real} random states, which are descriptive of the eigenstates of the MFIM. We begin by presenting the exact formulas for the mean and standard deviation of the EE distribution for complex random states ($\beta=2$). Then, we discuss how the asymptotic values are affected when the states are real $(\beta=1)$. 

\begin{figure}[b]
\centering\includegraphics[scale = 1.0]{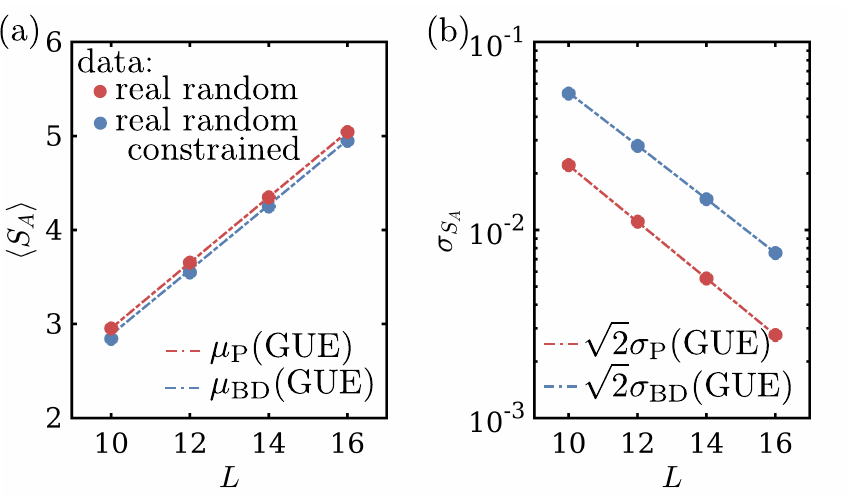}
  \caption{Numerical data for the mean entanglement entropy of real random states constrained to the $M=L/2$ symmetry sector (blue dots) and real random states in the full Hilbert space (red dots). The numerical data is obtained from $10^6$ randomly generated states.  For comparison, the analytic Page and BD means for the entanglement entropy distribution of random states with and without constraints are shown with dotted-dashed lines. (b) Numerical data for the standard deviations of the entanglement entropy using the same randomly generated data as in (a) (blue and red dots), compared against the analytic values of $\sigma_{\rm P}$ and $\sigma_{\rm BD}$ inflated by $\sqrt{2}$ to account for the larger standard deviation of real random states.} \label{fig:goedistribution}
\end{figure}

\subsubsection{Moments of the GUE ($\beta=2$)}

As shown in Ref.~\cite{2019PRD_BianchiDona}, the first moment of the distribution $P(S_A)$ for complex random states in a qubit system of size $L$ is given by 
\begin{align}
    \mu_{\rm P} &= \Psi(d_A d_B+1) - \Psi(d_B+1) - \frac{(d_A-1)}{2d_B}, 
\label{SAvgExact}
\end{align}
where $\Psi(x)=\Gamma'(x)/\Gamma(x)$ is the digamma function, defined as 
the logarithmic derivative of the Gamma function, and where $d_A=2^{L_A}$ , $d_B=2^{L-L_{A}}$ are the dimension of the subsystems $A$ and $B$. This expression can be evaluated exactly for all system and subsystem sizes. The asymptotic form of (\ref{SAvgExact}) in the limit of $L\rightarrow\infty $ and $f = L_A/L$ is written in Eq.(\ref{eq:Savg}) of the main text. The variance of the entanglement entropy distribution $\sigma_{\rm P}^2= \langle S_A^2\rangle -\langle S_A\rangle^2$ is given by
\begin{align}
    \sigma_{\rm P}^2&= \frac{d_A+d_B}{d_Ad_B+1}\Psi'(d_B+1) - \Psi'(d_Ad_B+1) \notag\\
    &\>\>\>\> - \frac{(d_A-1)(d_A+2d_B-1)}{4d_B^2(d_Ad_B+1)}. \label{SVarExact}
\end{align}
The asymptotic expression of Eq.(\ref{SVarExact}) in the limit of $L \rightarrow\infty$ and fixed $f=L_A/L$ is given in Eq.(\ref{eq:Svar}) of the main text. The numerical values for $\mu_{\rm P}$ and $\sigma_{\rm P}$ for different system sizes is shown in Table I. 

\begin{table}[]
    \vspace{.5cm}
    \centering
    \begin{tabular}{c||c|c||c|c|c|}
        $L$ & $\mu_{\rm P}$ & $\sigma_{\rm P}$  & $\mu_{\rm BD}$ & $\sigma_{\rm BD}$\\  
        \hline\hline
         8 & 2.2749 & 0.0311 & 2.2062 & 0.0718 \\
         \hline
         10 & 2.9663 & 0.0156 & 2.8866 & 0.0380 \\
        \hline
         12 & 3.6590 & 0.0078 & 3.5745 & 0.0199 \\
        \hline
         14 & 4.3521 & 0.0039 & 4.2652 & 0.0103\\
        \hline
         16 & 5.0452 & 0.0020 & 4.9569 & 0.0053\\       
    \end{tabular}
    \caption{Mean and standard deviation of the Page (P) and Bianchi-Dona (BD) distributions as a function of system size {for random complex states ($\beta=2)$} and $L_A = L/2$. For the BD distribution, we consider systems at half-filling, $m = M/L = 1/2$.}
    \label{tab:my_label}
\end{table}

\subsubsection{Moments of the GOE ($\beta=1$)}

Systems with TRS have a smaller effective Hilbert space dimension than those without TRS. Consequently, we expect the average entropy for real random states to be upper bounded by that of complex ones. While the exact expressions for the means differ, it was shown in Refs.~\cite{2010JPA_eermt,2011JPA_rmtee,2016PRE_entanglementdispersion} that they are asymptotically the same in the thermodynamic limit. However, the standard deviation acquires an additional $\sqrt{2}$ prefactor relative to the GUE ensemble, $\sigma_P^2(\beta = 1) \sim {2} \sigma_P^2(\beta = 2)$ since the GUE ensemble averages over both the real and complex parts. This behavior is explicitly shown in Fig.\ref{fig:goedistribution}, where the first two moments of $P_{\rm R}(S_A)$ are computed numerically by generating $10^6$ real random states, and compared against the asymptotic predictions, $\mu_P$ and $\sqrt{2}\sigma_P$. 

\subsection{The Bianchi Dona for constrained random states}

The presence of a U(1) charge causes the reduced density matrix in \eqref{rhoA} to become block diagonal in the charge sectors. To analyze the statistical properties of such systems, we present the exact formulas for the first two moments of complex random states subject to an additive constraint, which were first derived in Ref.~\cite{2019PRD_BianchiDona}. To the best of our knowledge, there are no known analytic results for constrained real states. However, based on numerical experiments, we show below that the conclusions of the unconstrained GOE ensemble translate to the constrained U(1) scenario, namely, fluctuations are enhanced by a factor of $\sqrt{2}$ relative to the GUE ensemble, whereas the means agree in the thermodynamic limit. 

\subsubsection{The constrained GUE ensemble ($\beta=2$)}

When the pure complex random states are constrained to a given symmetry sector $M$, the first moment of the distribution is given by \cite{2019PRD_BianchiDona}
\begin{align}
     \mu_{\rm BD}(M) = \sum_{M_A=0}^M & \frac{d_{M_A}}{d_M}(
     \mu_{\rm P}(M_A)+ \Psi(d_M+1)  \notag\\ & -\Psi(d_{M_A}+1)), \label{MeanSAFixedM}
\end{align}
where $\mu_{\rm P}(M_A)$ is the Page mean, Eqs.\eqref{SAvgExact}, constrained to the subspace ${\cal H}_A(M_A)\otimes{\cal H}_B(M-M_A)$, which means that the effective subsystem Hilbert space dimensions are $d_{A,M_A} = \binom{L_A}{M_A}$, $d_{B,M-M_A}= \binom{L-L_A}{M-M_A}, d_{M_A} = d_{A,M_A}d_{B,M-M_A}$, and the total Hilbert space dimension $\sum_{M_A} d_{A,M_A} d_{B,M-M_A} =d_M =\binom{L}{M}$. The asymptotic form of Eq.(\ref{MeanSAFixedM}) in the $L\rightarrow\infty$ for fixed $f=L_A/L$ and $m = M/L$ is given in Eq.(\ref{eq:SMMeanscaling}) of the main text. The second moment of the entanglement entropy distribution for random states constrained to the $M$-symmetry sector is given by 
\begin{widetext}
\begin{eqnarray}
\sigma_{\rm BD}^2= \sum_{M_A} \frac{d_{M_A}(d_{M_A}+1)}{d_M(d_M+1)} (( 
\sigma_{\rm P}^2(M_A)-\Psi'(d_M+2)+\Psi'(d_{M_A}+2)+(
\mu_{\rm P}(M_A)+\Psi(d_M+2)-\Psi(d_{M_A}+2))^2) \notag \\
+ \hspace{-.2cm}\sum_{M_{A}'\neq M_A} \hspace{-.2cm}\frac{d_{M_A}d_{M_A'}}{d_M(d_M+1)}((
\mu_{\rm P}(M_A')+ \Psi(d_M+2) -\Psi(d_{M_A'}+1))(
\mu_{\rm P}(M_A)+\Psi(d_M+2)-\Psi(d_{M_A}+1)) \notag \\ -\Psi'(d_M+2))- \mu_{\rm BD}^2,\,\,\label{VarSAFixedM}
\end{eqnarray}
\end{widetext}
where $\sigma_{\rm P}^2(M_A)$ is the Page variance, Eq.\eqref{SVarExact}, constrained to the subspace ${\cal H}_A(M_A)\otimes{\cal H}_B(M-M_A)$. The numerical values of $\mu_{\rm BD}$ and $\sigma_{\rm BD}$ for different system sizes is shown in Table I.

\subsubsection{The constrained GOE ensemble ($\beta=1$)}

Eigenstates of systems with TRS are real valued, thus it is necessary to extend Eqs.(\ref{MeanSAFixedM})-(\ref{VarSAFixedM}) for the case of real random states. We are not aware of any analytical results in this case. For this reason, we numerically compute the distribution of entanglement entropies of real random states constrained to the $M=L/2$ symmetry sector of a spin-1/2 chain. The moments are obtained by considering $10^6$ samples for system sizes ranging from $L=10$ to $L=16$. As shown in Fig.~\ref{fig:goedistribution}, the mean of the EE distribution agrees with analytic expression derived for complex states,  whereas the standard deviation increases by a factor of $\sqrt{2}$. This behavior is analogous to the analytical results obtained for unconstrained Haar random states. 

\begin{figure}[]
\centering
 \centering\includegraphics[scale = 1.0]{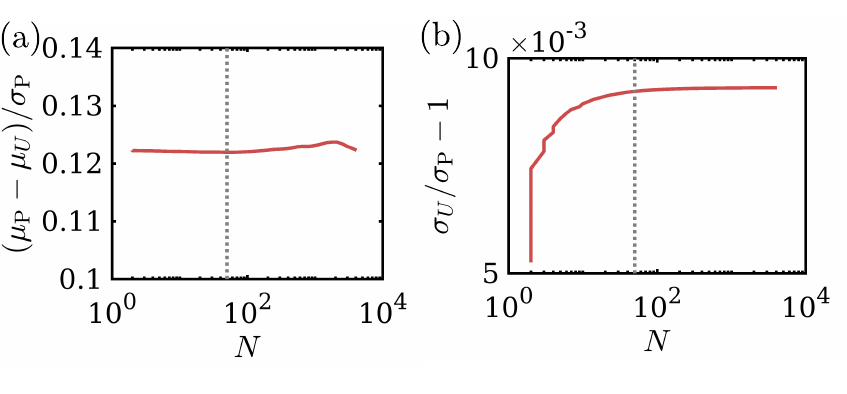}
  \caption{Average (a) mean and (b) standard deviation of the EE as a function of window size. For each value of $N$, we average over different centers for the window and over 100 circuit realizations. The vertical dotted lines indicates the window of 50 eigenstates used in the data of the main text. }
  \label{fig:windowSizeFRQC}
\end{figure}

\section{Window size dependence of the EE moments}
\label{app:windowsize}

In the main text, we discuss the number of eigenstates used to compute the microcanonical mean and standard deviations for the Hamiltonian and Floquet models. Here, we present numerical evidence that are results are not sensitive to the choice of window size, as long as the window is sufficiently large to get reliable estimates of the moments, and (for the Hamiltonian case) sufficiently small to avoid the effects of lower temperature eigenstates.  

\subsection{Floquet Random Circuits (FRC)}

In Figure \ref{fig:windowSizeFRQC}, we show data for the mean and standard deviation of $S_A$ for eigenstates of the FRQC on $L=12$ qubits as a function of window size $N$. For a given circuit realization, we divide the spectrum into groups of $N$ states closest in quasienergy. We compute the mean and standard deviation in each window of size $N$, then average these across the spectrum, and finally across circuit realizations. The first moment is independent of $N$ at this resolution observe that only approximately 50 eigenstates of the FRC are necessary for the ratios involving the first two moments to converge. In our data, we use the full spectrum for $L \leq 14$, while we use windows of 50 states centered at different quasienergies for $L=16$. 

 \begin{figure}[]
 \centering\includegraphics[scale = 1.0]{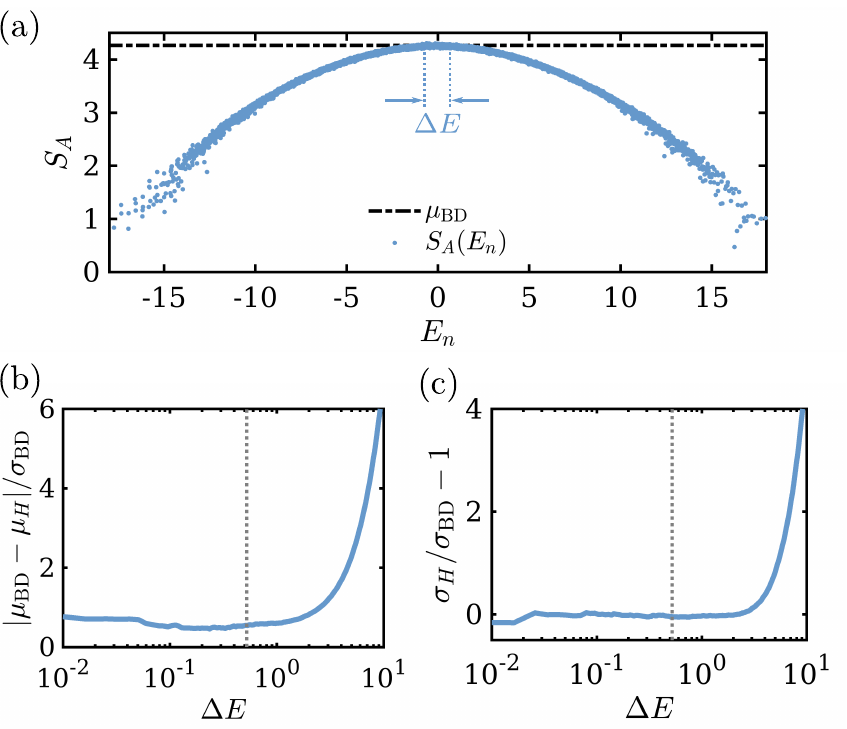}
   \caption{(a) Eigenstate entanglement entropy of the MFIM for model parameters $g = 1.08$, $h = 0.3$ and $L=14$. The parameter $\Delta E$ quantifies the energy window from which the variance of $S_A$ is computed. (b) Mean and (c) standard deviation of the entanglement entropy distribution as a function of the energy window width $\Delta E$. Shown with vertical dotted lines is the $\Delta E$ corresponding to a window of 600 eigenstates used in the main text.}
   \label{fig:dSdE}
 \end{figure}

\subsection{Mixed Field Ising Model (MFIM)}

When computing the microcanonical mean and variance of the distribution $P_{H}(S_A)$ for midspectrum energy density $E$ eigenstates, it is necessary to take a finite window $\Delta E$ in which to take samples of $S_A$.  In general, if $\Delta E$ is too small, then a statistically small number of states will be available for sampling thus resulting in large error bars. On the other hand, if $\Delta E$ is too large, then  low entanglement eigenstates will skew the distribution and increase its variance. Similarly to the FRC case,  we argue that due to typicality only few eigenstates are necessary to quantify the mean and standard deviation of the distribution. Here we numerically show that this is the case. In particular, we show that the value of the standard deviation of $S_A$ for midspectrum eigenstates does not vary appreciably when the window size is reasonably small, thus the results does discussed in the main text are quite insensitive to the choice of energy window width. 

In Fig.\ref{fig:dSdE}(a), we show the distribution of entanglement entropy computed for the MFIM with the MC  parameters $g = 1.08$ and $h=0.3$ for $L=14$. We compute the mean and variance of the distribution of $S_A$ computed for midspectrum eigenstates and using a variable window $\Delta E$ ranging from $\Delta E = 10^{-2}$ (approximately the typical eigenstate-to-eigenstate energy difference) to $\Delta E=10$ (approximately half the bandwidth of the system). We find that the mean and variance of the distribution does not vary significantly if $\Delta E \lesssim 2$. In the main text, we employ a total of 600 eigenstates for $L=14$  (see vertical dotted lines), which is a tiny fraction of the total number of states $2^{14} =16384$. We repeat the same analysis for all system sizes to define the width of the windows used in the main text for all system sizes $L$.

\section{Fluctuations of EE as a function of FRC gate range and period}
\label{app:gaterange}

In Fig.~\ref{fig:FRC} of the main text we found deviations from RMT in the second moment of the EE distribution of eigenstates that persist with increasing $L$. These differences were attributed to locality, the only feature present in the FRC. We now generalize the FRC model discussed in Sec.~\ref{sec:floquet} by relaxing the constraint of locality in order to observe the convergence to RMT. This will be achieved (a) by increasing the range of the local gates and (b) by increasing the number of periods of the FRC. As such, the circuit structure will be labeled by two parameters: the range $r$ and the period $T$, discussed in turn below. 

\begin{figure}[]
\centering\includegraphics[scale = 1.0]{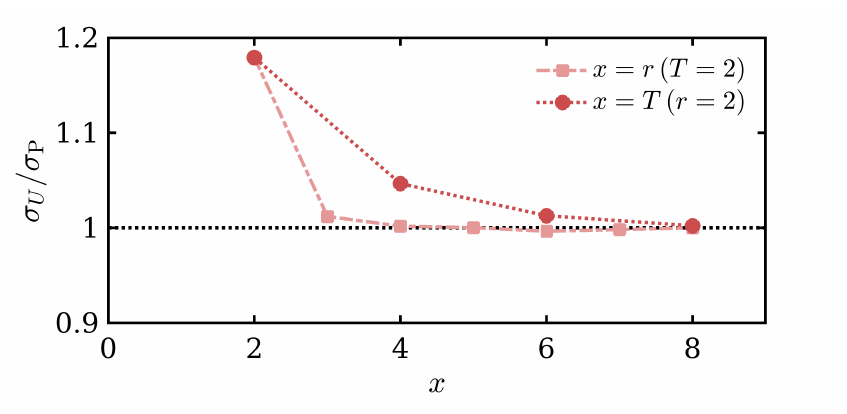}
  \caption{Convergence of the EE standard deviation of FRC eigenstates, $\sigma_U$, to the RMT prediction as a function of gate range $r$ for fixed circuit depth $T$ (squares), and as a function of circuit depth $T$ for fixed gate range (circles). Here we use a system of $L=16$ qubits. }
  \label{fig:gaterange}
\end{figure}

We consider brickwork circuits with staggered layers of range-$r$ unitary gates acting on a periodic one-dimensional spin-1/2 chain of length $L$. The range $r$ is the number of contiguous qubits each individual gate acts on, so that $U_{j, j+1, \cdots ,j+(r-1)}$ acts on sites $(j, j+1, \cdots ,j+r-1)$. Thus, $r=2$ denotes nearest-neighbour gates while $r=3$ is a three-site gate including both nearest and next-nearest neighbour interactions. The matrix $U_{j, j+1, \cdots ,j+(r-1)}$ is a random $U(n)$ matrix, with $n=2^r$. 

The generalized circuit architecture has a periodic brickwork layout with variable period $T \in \mathbb{Z}$. The circuit implements discrete time evolution, and advancing by one unit of time comprises the application of a ``layer'' comprised of $r$ staggered sub-layers. Each sub-layer is displaced by one lattice site with respect to the prior sublayer. For example, for the $r=2$ considered in the main text, advancing by one unit of time entails applying one layer of even and odd gates: 
\begin{equation}
    U(t+1,t) = \underbrace{\prod_j U_{2j+1, 2j+2}(t)}_{U_{\rm odd}(t) \equiv U_1(t)}  \underbrace{\prod_i U_{2i, 2i+1}(t)}_{U_{\rm even}(t) \equiv U_0(t)} .
\end{equation}
Likewise, $r=3$ requires applying three staggered sub-layers of gates starting from the $(0,1,2)$, $(1,2,3)$ and $(2,3,4)$ bonds respectively:
\begin{align}
    U(t+1,t) = &\underbrace{\prod_k U_{3k+2, 3k+3, 3k+4}(t)}_{U_2(t)} \times \nonumber \\  
    &\underbrace{\prod_j U_{3j+1, 3j+2, 3j+3}(t)}_{ U_1(t)} \times \nonumber \\ 
    &\underbrace{\prod_i U_{3i, 3i+1, 3i+2}(t)}_{U_0(t)}.
\end{align}
More generally, for range $r$ gates, 
\begin{align}
     U(t+1,t) &= \prod_{\alpha = 0}^{r-1} U_\alpha(t), \nonumber \\
     U_\alpha(t) &= \prod_j U_{rj + \alpha, rj + \alpha+ 1, \cdots, rj + \alpha+ r-1 }.
\end{align}
{In cases where $L$ is not divisible by $r$, we act with an identity matrix on the remaining sites.} For a circuit with periodicity $T$, the gates in the first $T$ layers are chosen independently,  and layers repeat after $T$ time-steps: $U(t+T+1,t+T) = U(t+1,t)$. The generalized Floquet unitary is defined as the time-evolution operator for period $T$:
\begin{equation}
    U_G(r,T) = \prod_{t=0}^{T-1} U(t+1,t),
\end{equation}
and $U(t=nT, 0) = U_G(r,T)^n$.

Figure~\ref{fig:gaterange} shows that the second moment of the EE distribution of eigenstates, $\sigma_U$, converges towards the RMT value $\sigma_{\rm P}$ both by increasing the period of the FRC (varying $T$ while fixing $r=2$), or by increasing the range of the gates (varying $r$ while fixing $T=2$).  These results corroborate the expectation that reducing locality increases convergence to RMT. However, it is striking that just increasing the gate range from $r=2$ to $r=3$ is already enough to remove most of the observed difference in standard deviation, suggesting that other (more sensitive) metrics might be needed to probe deviations from RMT.

\begin{figure}[]
\centering\includegraphics[scale = 1.0]{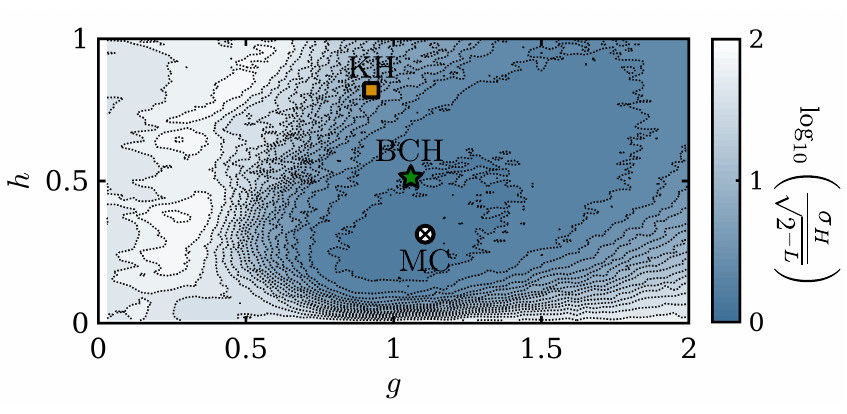}
  \caption{Colormaps of midspectrum entanglement entropy fluctuations normalized with $\sigma_{\rm BD}$, computed for the MFIM with transverse field $g$ and longitudinal field $h$, for $L=14$. We note that the colormap agrees qualitatively with the conclusions obtained form $D_{\rm KL}$ in Fig.\ref{fig:DKL}(a) of the main text.}
  \label{fig:sigmaratio}
\end{figure}

\section{Fluctuations of EE as a function of Hamiltonian model parameters}
\label{app:sigmaratio}

Since in most of the parameter space away from maximally chaotic (MC) point,  the main contribution to $D_{\rm KL}$ comes from the first moment which refers to the choice of the BD distribution, we now also present an alternate but related diagnostic of chaos which is agnostic to the reference RMT distribution, and only looks at the (normalized) fluctuations of EE: $\sigma_H/\sqrt{2^{-L}}$.  
This ratio is expected to be system-size independent and minimized for maximally chaotic systems, while being exponentially increasing for $L$ for near integrable systems. 
In this way, maximally chaotic Hamiltonians can be identified by minimizing the value of $\sigma_H$, even if the correct reference distribution is unknown. 
 In Fig.~\ref{fig:sigmaratio}, we show the normalized standard deviation of the microcanonical fluctuations of EE of eigenstates as a function of $(g,h)$. Similarly to Fig.~\ref{fig:DKL}(a) of the main text which accounts for both moments of the EE distribution, we find that $\sigma_H$ reaches a global minimum at the MC parameters and shows qualitatively similar behavior as $D_{\rm KL}$ away from the MC point. The colormap appears more noisy than Fig.\ref{fig:DKL}(a) because second moments have larger statistical fluctuations than first moments.

\begin{figure}
\centering\includegraphics[scale = 1.0]{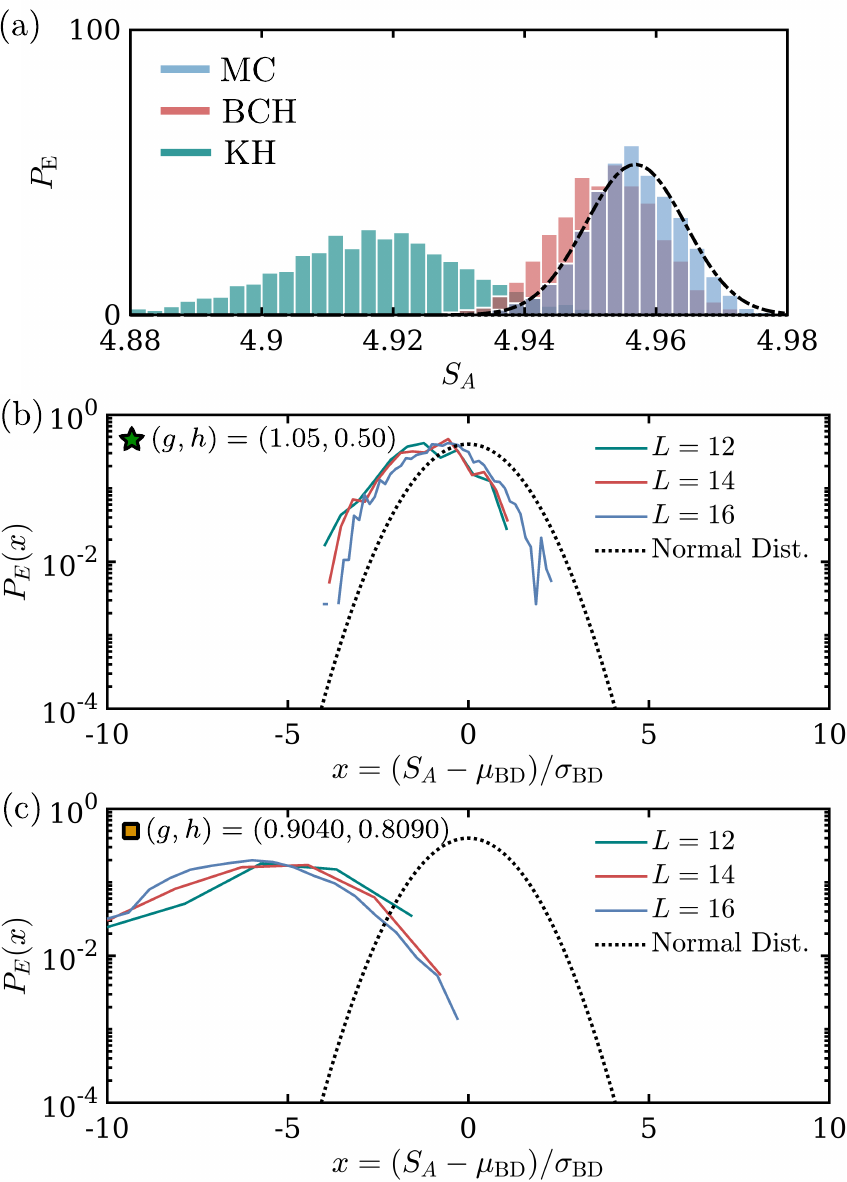}
\caption{(a) Distribution of entanglement for typically-used values of the MFIM: MC parameters in the present work (blue), the Banuls-Cirac-Hastings model (red), and Kim-Huse (green). Shown with dotted dashed line is BD distribution. Histogram of midspectrum EE obtained for the (b) BCH and (c) KH parameters as a function of system size, showing deviations from  RMT behavior that persist in the thermodynamic limit.  Shown with dotted lines is the reference standard normal distribution.}
\label{fig:typical_parameters}
\end{figure}

\section{Distribution of EE for some standard parameter values of the MFIM}
\label{app:MFIM_litcompare}

The MFIM is a paradigmatic model of strongly quantum chaotic system and has been routinely used in the study of quantum thermalization. In particular, there are several standard set of model parameters which are believed to be strongly chaotic. In this section, we compare the distribution of entanglement entropy for the most chaotic point found in the main text and previously-used parameters in the literature. Figure\,\ref{fig:typical_parameters}(a) shows histograms for the entanglement entropy of midspectrum eigenstates for (i) the most chaotic (MC) parameters of Fig.\ref{fig:DKL}, $(g,h) = (1.08,0.30)$ (blue bars), (ii) for the BCH model\cite{BanulsCiracHastings} (green bars), $(g,h)=(1.05,-0.5)$ and (iii) for the Kim-Huse model $(g,h)=((\sqrt{5}+5)/8 ,(\sqrt{5}+1)/4) \approx (0.9045,0.8090)$ \cite{Kim_2013} (red bars). We find that the BCH parameters agree reasonably well with the BD distribution, whereas the KH parameter strongly deviates by more than two standard deviations. 

A more refined look into the distribution of entanglement entropy normalized with the BD distribution is shown in Fig.\ref{fig:typical_parameters}(b-c). We observe that, within the scale of $\sigma_{\rm BD}$, eigenstates in the BCH model agree well with RMT behavior: the means of the EE distributions differ by  $\Delta\mu \sim\sigma_{\rm BD}$. In contrast, for the KH parameters, we observe large deviations between  the EE distribution of eigenstates and random states: in this case, the means differ by $\Delta\mu \sim 5\sigma_{\rm BD}$, and the standard deviations is around five times larger.

\end{document}